\documentclass[aps,prb,showpacs,amsmath,amssymb,twocolumn]{revtex4}
\pdfoutput=1

\usepackage{xcolor}
\usepackage{graphicx}
\usepackage{amsmath}
\usepackage{amssymb}
\usepackage{amsfonts}
\usepackage{units}
\usepackage{bm}
\usepackage{hyperref}
\usepackage{enumerate}

\usepackage{longtable}
\usepackage{totcount}

\begin{document}

\title{Nonlinear electromagnetic response and Higgs mode excitation in BCS superconductors with impurities}

\date{\today}

\author{Mikhail Silaev}
 \affiliation{Department of
Physics and Nanoscience Center, University of Jyv\"askyl\"a, P.O.
Box 35 (YFL), FI-40014 University of Jyv\"askyl\"a, Finland}

\begin{abstract}
We reveal that due to the presence of disorder oscillations of the order parameter amplitude called the Higgs mode can be effectively excited by the external electromagnetic radiation in usual BCS superconductors.  
 This mechanism works for superconductors with both isotropic s-wave and anisotropic, such as d-wave, pairings.  
The non-linear response in the presence of impurities is captured by the quasiclassical formalism. We demonstrate that analytical solutions of the Eilenberger equation with impurity collision integral and external field drive coincide with the exact summation of ladder impurity diagrams.
 Using the developed formalism we show that  resonant third-harmonic signal observed in recent experiments is naturally explained by the excitation of Higgs mode mediated by impurity scattering. 
 \end{abstract}
 
 \maketitle
 %%%%%%%%%%%%%%%%%%%%%%%%%%%%%%%%%%%%%%%%%%%%%%%%%%%%%%%%%%
 \section{Introduction}
 Nonlinear electromagnetic responses are ubiquitous in superconducting 
 systems and have attracted interest for many years 
 \cite{GorkovTDGL1968, Gorkov1968,PhysRevLett.37.930,Gorkov1969}. 
 External field with the frequency $\Omega$ produces several 
 important second-order corrections to the superconducting
  order parameter $\Delta$. 
 The  zero-frequency change of $\Delta$ leads to the critical temperature and critical current enhancements  
 \cite{Ivlev1971, Eliashberg1970,PhysRevB.97.184516} known as the microwave stimulation of the superconducting state. 
 The time-dependent corrections to $\Delta$ at the frequency $2\Omega$ produce the current oscillating with frequency \cite{GorkovTDGL1968} $3\Omega$. 
 This effect called the third harmonic generation (THG)  has been observed 
 experimentally in microwaves and explained with the help of the time-dependent Ginzburg-Landau theory
 \cite{PhysRevLett.37.930}.

 Recently the terahertz (THz) spectroscopy of superconducting state has  
 become experimentally available \cite{PhysRevLett.107.177007, 
 PhysRevLett.110.267003, Matsunaga2012,Matsunda2013,Matsunaga1145,Giorgianni2019}. 
 This range of frequencies is especially interesting since it overlaps with  
 the typical gap sizes in low-temperature superconductors like NbN. 
 Thus measuring nonlinear responses in THz domain allows 
 for probing dynamics of the order parameter amplitude 
 \cite{Matsunda2013,Matsunaga1145} predicted to feature oscillations 
 with an eigen frequency $2\Delta (T)$ where 
 $\Delta (T)$ is the the gap at a given temperature \cite{Volkov1973,  
 Kulik1981,Barankov2004,Barankov2006}. 
 By analogy with the Higgs boson\cite{Higgs1964} in particle physics 
 this type of collective excitation in condensed matter systems is 
 called the Higgs mode\cite{Varma2002,Podolsky2011,Pashkin2014,Volovik2014,Varma2015}.
 The order parameter amplitude oscillations excited by the short optical 
 pulses has been observed in several pump-probe experiments 
 \cite{Matsunda2013,Matsunaga1145}. 
 Recent measurements report the evidence of resonant Higgs mode 
 excitation in the THG component of the THz signal transmitted 
 through the superconducting plate\cite{Matsunaga1145,PhysRevB.96.020505}.
 Earlier the amplitude  modes have been  
 observed by Raman scattering in superconductors with charge density wave order \cite{Sooryakumar1980,PhysRevB.26.4883, PhysRevB.97.094502, 1806.03433} and by the nuclear magnetic resonance in superfluid $^3$He \cite{PhysRevLett.30.829,PhysRevLett.30.541, Zavjalov2016}.
  %
  %%%%%%%%%%%%%%%%%%%%%%%%%%%%%%%%%%%%%%%%%%%%%%%
 \begin{figure}[htb!]
 \centerline{$
 \begin{array}{c} 
 \includegraphics[width=0.8\linewidth]{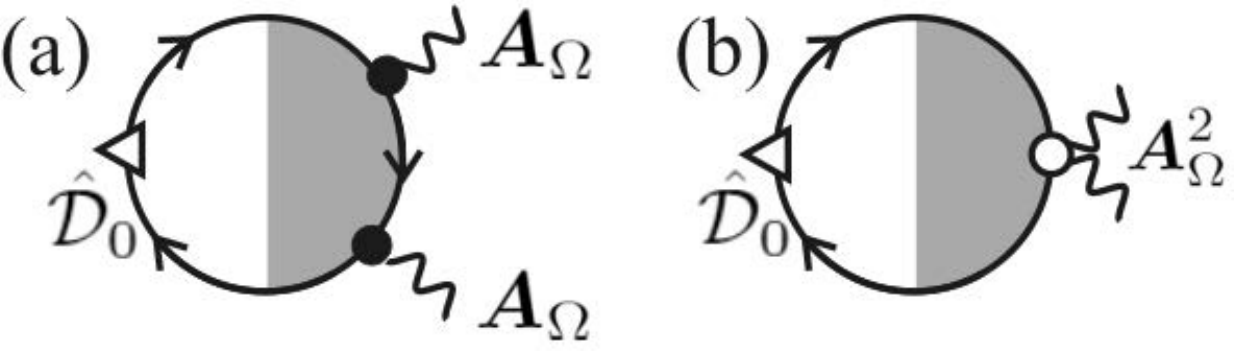}  
  \end{array}$}
 \caption{\label{Fig:Higgs} (Color online) 
 Two possible processes contributing to the order parameter 
 excitation at the double frequency of external field described by 
 the vector potential $\bm A_\Omega$. 
 (a) The second-order diagram with two current vertices $\bullet$
 corresponding to the light-matter coupling linear by the external field $\bm A_\Omega$.  
 The impurity ladder insertion is shown by the shaded gray area.
  (b) The first-order diagram with momentum-independent density vertex.  
  The order parameter vertex ${\cal D}_0$ is defined below in Eq.(\ref{Eq:D0}).
 }
 \end{figure}   
 %%%%%%%%%%%%%%%%%%%%%%%%%%%%%%%%%%%%%%%%%%%%5   
 
 %%%%%%%%%%%%%%%%%%%%%%%%
 Despite the significant experimental advances, 
 theoretical understanding of high-frequency non-linear properties in  
 superconductors is still lacking. Numerical simulations 
 presented in several works consider strongly non-equilibrium regimes  
 without any disorder 
 \cite{Krull2014,Papenkort2008, 1712.07989} and have to attribute rather large wavevector to the radiation field in order to obtain the sizable coupling with the order parameter. 
 At the same time perturbative calculations of  nonlinear responses reveal several important 
 limitations imposed by the absence of impurity scattering.
 The order parameter modulation by the external radiation was studied in the pioneering work of Gor'kov and Eliashberg
 \cite{Gorkov1968} who considered the electron-photon coupling 
 linear by the 
 vector potential $\bm A_\Omega$.
 The contribution of such terms to the  order parameter 
 modulation  $\Delta_{2\Omega}$ at the frequency $2\Omega$ is
 described by the second-order diagram in Fig.\ref{Fig:Higgs}a. 
 Here current vertices $\bullet$ describe the linear coupling to the external field. This contribution disappears  
 the absence of impurity scattering\cite{Gorkov1968}.  
 One can think of this as a consequence of the Galilean invariance featured by the superconducting condensate. Indeed switching to the moving frame can eliminate  the condensate velocity induced by external field. In the moving frame the order parameter amplitude remains unaffected thus coinciding with that of the stationary condensate. Therefore,  in order to perturb the amplitude through the linear electron-photon coupling terms  the Galilean invariance should be broken either by the spatially-inhomogeneous field or the inhomogeneous potential which can naturally related to the presence of impurities.  
  
 Based on these arguments at zero or negligibly small wave vectors the Higgs mode excitation \cite{PhysRevB.92.064508} and
 non-linear response\cite{Cea2016} in the absence of impurities are
 possible only through the electron density modulation generated by the term $\propto A_\Omega^2$ in the Hamiltonian. The corresponding density vertices $\bigcirc$ are  similar to those which determine Raman scattering in superconductors
 \cite{AbrikosovRaman1961, Abrikosov1974, Sooryakumar1980,
 Abrikosov1988,Falkovsky1993}. The perturbation of the order parameter due to such coupling 
 is shown by the diagram in Fig.\ref{Fig:Higgs}b.  
 However, this contribution to $\Delta_{2\Omega}$
 vanishes within the Bardeen-Cooper-Schrieffer (BCS) model of superconductivity and becomes non-zero only due to the various extensions \cite{PhysRevB.92.064508, Cea2016,PhysRevB.96.020505}. 
 
 The prediction of negligible Higgs mode generation in BCS superconductors  \cite{Cea2016} has been in contrast both to the recent
 THz probes\cite{PhysRevLett.110.267003, PhysRevLett.107.177007,Matsunaga1145,Matsunda2013,Matsunaga2012, PhysRevLett.120.117001,PhysRevB.96.020505} and earlier microwave experiments\cite{PhysRevLett.37.930} where significant 
 radiation-induced oscillation $\Delta_{2\Omega}$ have been observed. 
 %
 %%%%%%%%%%%%%%%%%%%%%%%%%%%%%%%%%%%%%%%%%%%%%% 
 Here we resolve this controversy and show that previous theoretical conclusions about the non-linear response of superconductors are drastically 
 altered in the presence of disorder. We consider the arbitrary amount of impurity 
 scattering treated within the self-consistent Born approximation. The 
 calculations are implemented using both the diagrammatic technique and 
 quasiclassical Eilenberger theory formalism \cite{Eilenberger1968} and the 
 agreement between these two approaches is demonstrated. 
 We show that in the presence of disorder the 
 non-linear process shown in Fig. \ref{Fig:Higgs}a  provides quite an effective generation of the Higgs mode which shows up through the resonant THG in agreement with several recent experiments.

 { In the diffusive limit typical for the superconducting thin films  our findings confirm that the resonant third-harmonic generation at the frequency $\Omega=\Delta(T)$ observed in \cite{Matsunaga1145} is determined solely by the Higgs mode generation. This result is in sharp contrast with the system without impurities where the Higgs mode generation is negligible and the resonant contribution comes from the other source \cite{Cea2016,Cea2018}. It agrees qualitatively with the
  studies of linear Higgs mode generation in diffusive current-carrying superconductor\cite{PhysRevLett.118.047001,1809.10335}. 
  }
         
 %%%%%%%%%%%%%%%%%%%%%%%%%%%%%%%%%%%%%%%%%%%%       
 \section{Formalism }
    
 \subsection{General approach}    
 %%%%%%%%%%%%%%%%%%%%%%%%%%%%%%%%%%%%%%%%%%%%%%%%
 We describe the interaction of electrons with electromagnetic field 
 using the following Hamiltonian which contains two qualitatively 
 different terms \cite{Abrikosov1974,Abrikosov1988,Falkovsky1993}
   \begin{align} \label{Eq:Hp}  
    & \hat H_p = \hat V_1 + \hat V_2 
    \\ \label{Eq:V1}
    & \hat  V_1 = -\frac{e}{c} ({\bm v \bm A}) 
    \\ \label{Eq:v2}
   & \hat V_2 = \hat\tau_3 \frac{e^2}{2mc^2} A^2,
   \end{align}      
   where $\bm v = \partial E_p/\partial \bm p$ is the band velocity, 
   $m$ and $e$ are the electron mass and charge. 
       
  Here we introduce the notation  $\hat\tau_{0,1,2,3}$ for the Pauli matrices in Nambu particle-hole space.   
 In the diagrammatic representation 
 the perturbation term  $\hat V_1$ linear in the external field amplitude is described by the current vertex with attached single external field line. Such current vertices determine diagrams of the type shown in Fig.\ref{Fig:Higgs}a. The term $\hat V_2$ quadratic by the external field produces radiation-induced electronic density modulation.  Thus light-matter coupling is described by  diagrams  with density vertices $\propto \hat\tau_3$ such as shown by the open circles in Fig.\ref{Fig:Higgs}b.
   
   The charge current and order parameter as functions of the at imaginary time $\tau [0, \beta]$ are given by  
   \begin{align}
   & \bm j (\tau) =  e\int \frac{d^3\bm p}{(2\pi)^3} 
   \left(\bm v - \tau_3 \frac{e }{c} 
   \hat m^{-1} \bm A (\tau) \right) \hat G (\bm p, \tau_{1,2}=\tau)  
   \\
  & \hat\Delta (\tau)= \tilde\lambda \int \frac{d^3\bm p}{(2\pi)^3} \hat P [\hat G] 
   (\bm p, \tau_{1,2}=\tau) 
   \end{align}
 where $\hat G (\bm p, \tau_{1,2})$ is the imaginary time 
 Green's function (GF), $ \tilde\lambda$  is the pairing constant,
 $\hat P [\hat G] = \hat\tau_1 {\rm Tr}(\hat\tau_1\hat G) +
 \hat\tau_2 {\rm Tr}(\hat\tau_2\hat G)$ is the  projection operator. 
 It will be convenient to use also the frequency  representation which can 
 be defined as 
 $\hat G (\omega, \Omega) = 
 \int d\tau_1 d\tau_2 \hat G(\tau_1,\tau_2) e^{i\omega (\tau_1-\tau_2) + 
 i\Omega (\tau_1+\tau_2)}$. In this representation the 
 self-consistency relation can be written as 
 $\Delta_{2\Omega} = \hat {\cal D}_0 [\hat G] $, where 
 $\hat {\cal D}_0$ is the order parameter vertex 
 which appears first in the diagrams in Fig.(\ref{Fig:Higgs}). 
 The algebraical expression for $\hat {\cal D}_0$ reads
 %%%%%%%%%%%%%%%%
 \begin{align} \label{Eq:D0}
 \hat {\cal D}_0 [\hat G]= \tilde\lambda T \sum_\omega \int \frac{d^3\bm p}
 {(2\pi)^3} \hat P [\hat G](\omega,\Omega,\bm p).
 \end{align}          
 %%%%%%%%%%%%%%%%
 The stationary propagators depend only on the frequency corresponding to 
 the relative time $\hat G = \hat G_0(\omega, \bm p)$. 
 In disordered superconductors  with arbitrary amount of point-like 
 impurity scatterers treated within the self-consistent Born approximation  
 propagators are given by \citep{AGD} 
 %%%%%%%%%%%%%%%%%%%%%%%%%%%%%%%%%%%%%%%%%%%%%%%%
 \begin{align} \label{Eq:GFimp}
 & \hat G_0 (\omega,\bm p)= \frac{ \tilde\Delta \hat\tau_1 -
 i\tilde{\omega} 
 \hat\tau_0 - \xi_p \hat\tau_3}
 {\tilde\Delta^2 + \tilde{\omega}^2 + \xi_p^2}
 \\
 %& \tilde{\omega}= \omega \left( 1 + \frac{1}{2\tau_{imp}
 %\sqrt{\omega^2 + \Delta^2}} \right) 
 & \tilde{\omega}= \omega \frac{\tilde{s} (\omega) }{s(\omega)};
 \;\; \tilde{\Delta}= \Delta \frac{\tilde{s} (\omega) }{s(\omega)} , 
 %\left( 1 + \frac{1}{2\tau_{imp}
 %\sqrt{\omega^2 + \Delta^2}} \right)
 \end{align}
 %%%
 where $\xi_p = p^2/2m - \mu$ is the deviation of the kinetic  
 energy from the chemical potential $\mu$ and 
 $\hat \tau_{1,2,3}$ are the Pauli matrices in Nambu space. 
 We denote $s= \sqrt{\omega^2 + \Delta^2}$ and $\tilde{s} = s + 1/2\tau_{imp}$.  
 
 The 
 propagator (\ref{Eq:GFimp}) is averaged over the randomly 
 disordered point scatterers configurations. It is parametrised 
 by the scattering time $\tau_{imp}^{-1} = 2\pi \nu n u^2 $, where 
 $\nu$ is the normal metal density of states at the Fermi level, 
 $u$ is the impurity potential strength and $n$
 is the density of impurities. 
        
\begin{figure}[htb!]
 \centerline{$
 \begin{array}{c}
 \includegraphics[width=0.98\linewidth]{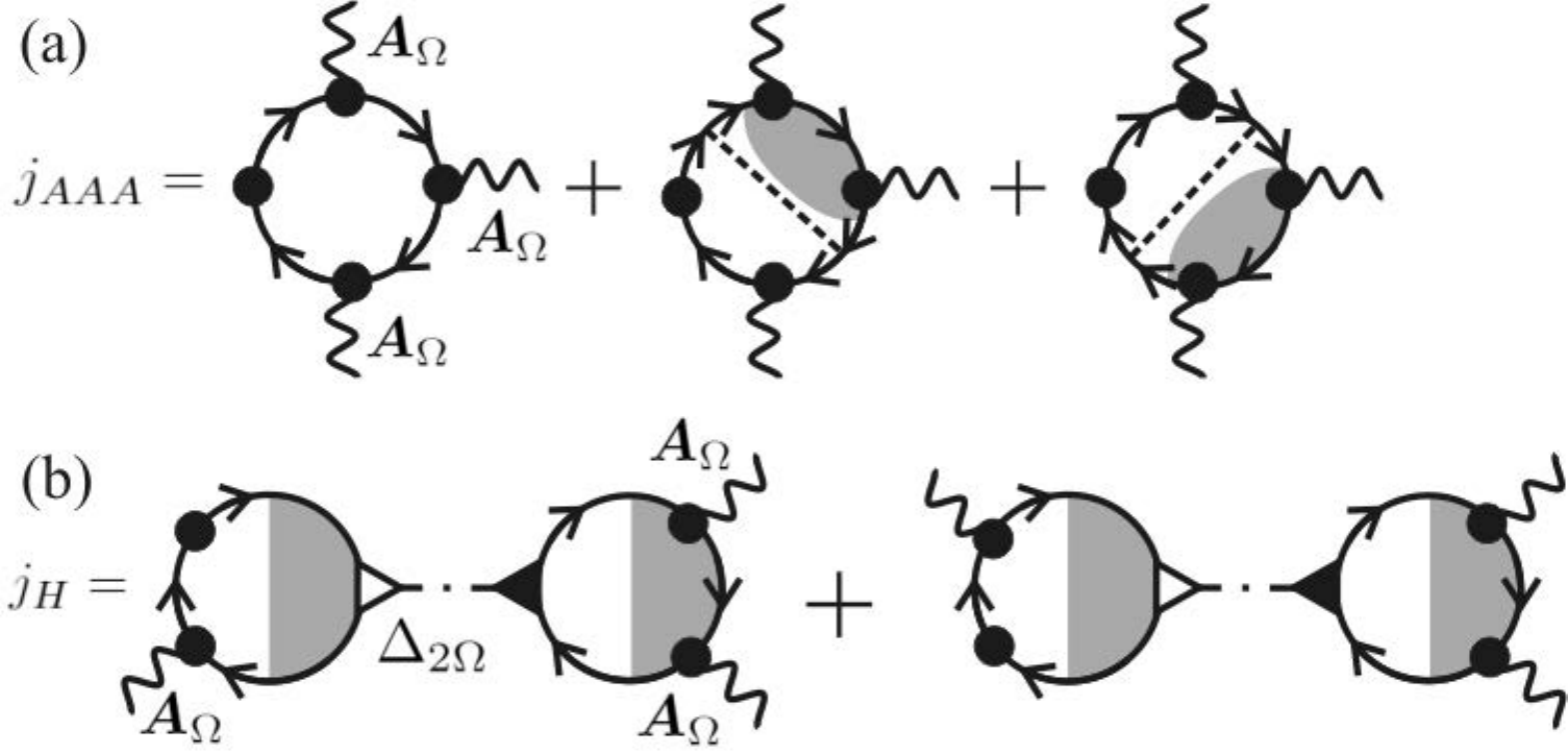}
 \end{array}$}
 \caption{\label{Fig:Paramagnetic} 
Diagrams with current vertices $\bullet = e \bm v/c $ contributing to non-linear response. (a) Direct third-order coupling to the external field $\bm A_\Omega$. 
 The dashed lines show impurity scattering correction  in Born approximation averaged by Gaussian disorder. Shaded regions show impurity ladders.  (b) Coupling to the external field through the excitation of the order parameter oscillation $\Delta_{2\Omega}$ shown by the dash-dotted line and the corresponding vertex is $\rhd = \hat \tau_1$. The  filled triangle corresponds to the order parameter vertex modified by the polarization bubble insertions which yield the Higgs mode excitation.   
 }
 \end{figure}

 We are interested in the non-linear contribution to the current 
 determined  by the diagrams  in 
 Fig.\ref{Fig:Paramagnetic} where the coupling to the 
 electromagnetic field is determined by the vector current-type 
 vertices. Such diagrams are generated by the perturbation 
 potential $\hat V_1$ given by Eq.(\ref{Eq:V1}).  
 As we see below the contribution of such diagrams to the 
 measurable quantities is captured by the quasiclassical 
 Eilenberger equations \cite{Eilenberger1968}. We assume here  
 that all external field lines have the same time dependence 
 $\bm A_\Omega e^{i\Omega t}$, 
 so that diagrams in Fig.\ref{Fig:Paramagnetic} yield the third 
 harmonic response of the current 
 $(j_{AAA} + j_H) e^{3i\Omega t}$.
  
 The diagrams shown in Fig.(\ref{Fig:Paramagnetic}a) determine 
 the current $j_{AAA}$ generated by the direct coupling to the 
 external field through the current vertices. Dashed lines 
 correspond to impurity scattering averaged over 
 the random disorder configuration. Analytically 
 one should integrate  over the input/output momentum the 
 content between the dashed line start and end points as well 
 as multiply the result by the factor $1/(2\pi \tau_{imp})$. 
 The shaded regions show impurity ladders discussed in detail 
 in Sec.\ref{Eq:NonLinearVertex}. 
   %
%  The second-order corrections due to this vertex are calculated in Sec.(\ref{Eq:NonLinearVertex}).

%%%%%%%%%%%%%%%%%%%%%%%%%%%%%%%%%%%%%%%%%%%%%%%
 %%%%%%%%%%%%%%%%%%%%%%%%%%%%%%%%%%%%%%%%%%%%%%%%%%%  
    
\begin{figure}[htb!]
  \centerline{$
  \begin{array}{c}
  \includegraphics[width=0.7\linewidth]{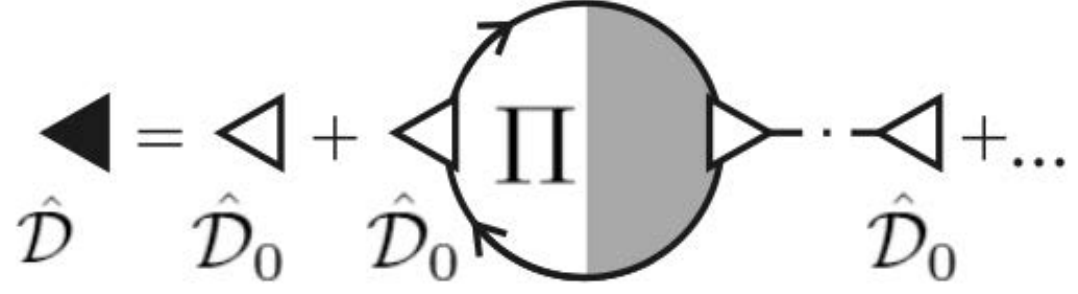}
  %\put (-300,80) {\LARGE (a) } %\put (-300,50) {\LARGE $j=$ }
  \end{array}$}
  \caption{\label{Fig:Polarization} 
  Order parameter vertex $\cal D$ (filled triangle) corrected by the 
  polarization operator insertions. $\rhd = \hat \tau_1$, $\lhd = \hat {\cal D}_0$, dashed region corresponds to the 
  impurity ladder insertion.   
  }
  \end{figure}

Besides the direct coupling to external field equally important is the third-harmonic generation  by the current $j_H$ determined by the order parameter modulation $\Delta_{2\Omega} e^{i2\Omega \tau}$. Taking into account that $\Delta_{2\Omega}$ is generated by external source according to the diagram in Fig.\ref{Fig:Higgs}a the current $j_H$ can be represented by the third-order response diagram in Fig.\ref{Fig:Paramagnetic}b. This contribution is of the special interest since as we will demonstrate its frequency dependence contains the information about the Higgs mode, that is the resonant enhanceable of the order parameter oscillations amplitude at $\Omega =\Delta$. Technically the resonant Higgs mode contribution is determined by the polarization operator which modifies the gap function vertex shown by the filled triangle in Fig.(\ref{Fig:Paramagnetic}b). This  vertex yields the order parameter coupling to the external source such as the second-order correction of the GF by the electromagnetic field discussed above. The diagrammatic representation of the  order parameter vertex with polarization bubble insertions is shown in Fig.(\ref{Fig:Polarization}). Note that it contains impurity scattering corrections, both as the self-energies modifying individual propagator lines and the ladder dressing the $\rhd = \hat \tau_1$ vertex.

 %%%%%%%%%%%%%%%%%%%%%%%%%%%%%%%%%%%%%%%%%%%%%%
 %%%%%%%%%%%%%%%%%%%%%%%%%%%%%%%%%%%%%%%%%%%%%%%%%%% 
            
\begin{figure}[htb!]
 \centerline{$
 \begin{array}{c}
 \includegraphics[width=0.6\linewidth]{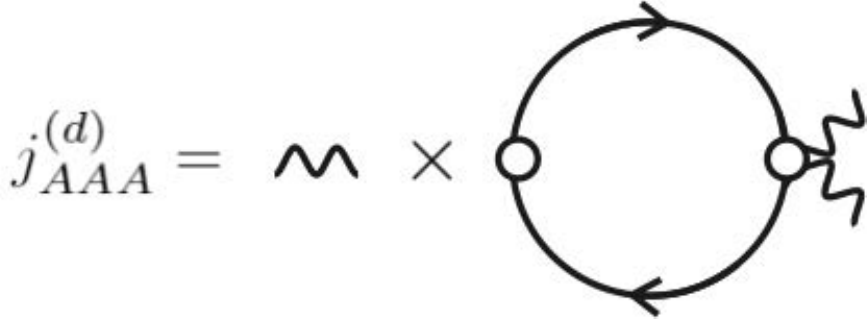}
 %\put (-200,30) {\LARGE (a) } \put (-170,30) 
 %{\LARGE $j^{(d)}_{AAA}=$ }
 \end{array}$}
 \caption{\label{Fig:Diamagnetic} 
 Diagram contributing to the 
 THG current response through the density modulation 
 due to the direct coupling to 
 the vector potential $\bm A_{\Omega}$ denoted by wavy lines.
 Density vertices $\bigcirc = \hat\tau_3 (e^2/2mc^2) $
 are same as in  Fig.\ref{Fig:Higgs}b.  
 }
 \end{figure}

 Previously, it has been shown that in the the absence of impurities 
 $\tau_{imp}=\infty$ 
 the contribution of diagrams with current vertices ${\Large{\bullet}}$ to the order parameter 
 modulation $\Delta_{2\Omega}$ disappears 
 \cite{Gorkov1968, PhysRevB.92.064508,Matsunaga1145, Cea2016}. 
 In this case the  only non-zero contribution is given by the 
 diagrams with density vertices $\bigcirc$ shown in Fig.\ref{Fig:Diamagnetic}. 
 Below we compare the contributions of these diagrams to obtain the 
 threshold impurity concentration when the paramagnetic diagrams 
 start to dominate.   

  %%%%%%%%%%%%%%%%%%%%%%%%%%%%%%%%%%%%%%%%%%%%%%%
 %%%%%%%%%%%%%%%%%%%%%%%%%%%%%%%%%%%%%%%%%%%%%%%%%%%  

 %%%%%%%%%%%%%%%%%%%%%%%%%%%%%%%%%%%%%%%%%%%  
 \subsection{Quasiclassical theory}        
 \label{Sec:QuasiclassicalFormalism}
    
 In general it is believed that the quasiclassical approximation introduced 
 by Eilenberger \cite{Eilenberger1968} takes into account all diagrams with 
 current vertices shown in Fig. (\ref{Fig:Paramagnetic}) but neglects the ones with 
 density vertices. Technically this happens because the 
 $\hat \tau_3 A^2$ terms in the Hamiltonian generically drop out from the
 quasiclassical equations. However, previously there has been  done no direct comparison of the results given by quasiclassical theory with those obtained from diagrams describing coupling to the external field in the presence of impurity scattering. 
 We will implement this check and demonstrate the summation of diagrams with impurity ladders give the same results as the quasiclassical calculation implemented according  to the formalism described below. 
         
     The quasiclassical propagator is defined as 
     \begin{align}\label{Eq:QuasiclassicalGF}
     \hat g= \frac{i}{\pi} \int d\xi_p \hat\tau_3\hat G
     \end{align}
     In the imaginary time domain $\hat g = \hat g (\tau_1,\tau_2,\bm r,\bm v_F)$
     is determined by the Eilenberger equation \cite{Eilenberger1968}
    \begin{align} \label{Eq:Eilenberger}
    & \frac{ie}{c}\bm v_F [\hat\tau_3\bm A,\hat g]_\tau  = 
    \\ \nonumber
    & - i\{ \hat
    \tau_3\partial_\tau , \hat g \}_\tau + 
    i[\hat\tau_3\hat\Delta, \hat g]_\tau + 
    \frac{1}{2\tau_{imp}}[\langle \hat g\rangle \circ, \hat g ].
    \end{align}  
    Here we denote the commutators 
    $[X,g]_\tau= X(\tau_1) g(\tau_1,\tau_2) - 
    g(\tau_1,\tau_2)X(\tau_2)$ 
    and the convolution is given by 
    $\langle \hat g\rangle\circ \hat g  = 
    \int_0^\beta d\tau \langle \hat g\rangle(\tau_1,\tau) 
    \hat g (\tau,\tau_2) $. 
    The angle-averaging over the Fermi surface is given by 
    $\langle g\rangle$. 
      The current and order parameter are given by 
      %%%%%%%%%%%555      
     \begin{align} \label{Eq:CurrentQuasiclassic}
     & \bm j (\tau) = -i \pi e \nu {\rm Tr} 
     [\hat\tau_3 \langle \bm v_F \hat g (\tau,\tau) \rangle ]
      \\ \label{Eq:OPQuasiclassic}
     & \hat \Delta (\tau) = -i  \lambda 
     \hat P [ \hat\tau_3\langle \hat g (\tau,\tau) \rangle ],
      \end{align}
 %%%%%%%%%%%%%%%%%%%%%%%%
 where $\lambda = \pi\nu \tilde{\lambda}$ and $\bm v_F = \bm v (\bm p=\bm p_F)$ is the Fermi velocity which is  the band velocity 
 at the momentum equal to the Fermi momentum $\bm p_F$.
 The quasiclassical equations are supplemented by the normalization 
 condition $ \hat g\circ \hat g =1$.
      
 %%%%%%%%%%%%%%%%%%%%%%%%%%%%%%%%%%%%%%%%%%% 
 {\it Dirty limit: Usadel theory.}
 In the dirty limit $\tau_{imp}T_c \ll 1$ the calculations can be 
 significantly simplified using the Usadel equation formalism\cite{PhysRevLett.25.507}.
 The key idea of this approximation is to represent the GF as the superposition of isotropic $\langle \hat g\rangle$ and  anisotropic $\hat g_{an} \propto (\bm v_F\bm A)$ parts.
  For the validity of Usadel theory it  
 is required that the anisotropic part is small 
 $\hat g_{an}\ll \langle \hat g\rangle$ which is
satisfied in the diffusive limit. 
  Then using the normalization condition 
 $\langle \hat g\rangle\circ \hat g_{an} + 
  \hat g_{an}\circ\langle \hat g\rangle =0$ 
  one obtain  get the 
  Usadel equation for the isotropic component $\langle \hat g\rangle$ which reads  (we omit angular brackets)
 %%%%%
 \begin{align} \label{Eq:UsadelGen}
 & -i \{\hat\tau_3\partial_\tau, \hat g \}_\tau + i[\hat\tau_3 \hat\Delta , 
 \hat g ]_\tau =  \hat\Sigma\circ\hat g - \hat g\circ\hat\Sigma
 \\ \label{Eq:MWCI}
 & \hat\Sigma_{em} (\tau_1,\tau_2) = \frac{D e^2}{c^2}\bm A(\tau_1)\hat\tau_3
 \hat g (\tau_1,\tau_2)
 \hat\tau_3 \bm A (\tau_2) 
 \\ \label{Eq:CurrentUsadel}
 & \bm j (\tau) = 
  \pi\frac{\sigma}{c} \times
  \\ \nonumber
 &{\rm Tr}[ \hat\tau_3 
 \hat g(\tau,\tau_1)\circ 
 \left(
 \hat g(\tau_1,\tau)\bm A (\tau)\hat\tau_3
 -
 \hat\tau_3\bm A(\tau_1) \hat g(\tau_1,\tau) 
 \right) ]
 \end{align}
 %%%%%%
 where the self-energy $\hat\Sigma_{em}$ describes coupling to the electromagnetic field,  the  the diffusion coefficient is $D=\tau_{imp} v_F^2/3$ and 
 conductivity is $\sigma = e^2 \nu D$.

% The convolution operator is given by 
% $A\circ B= \int_0^\beta d\tau  A(\tau_1,\tau)B(\tau,\tau_2)$ 
% and 
% $[\hat\tau_3\hat\Delta, \hat g  ]_\tau = \hat\tau_3\hat\Delta(\tau_1)\hat 
% g(\tau_1,\tau_2) - 
% \hat g(\tau_1,\tau_2)\hat\tau_3\hat\Delta(\tau_2)$.        

 \section{Quasiclassical calculations}          
 First, we obtain time-dependent perturbations of the order parameter 
 and the third-order current response using the Eilenberger 
 theory for quasiclassical propagators. In the Sec.\ref{Sec:DiagramSummation} these
 results will be confirmed by the direct summation of  impurity ladder diagrams with 
 current-type vertices describing the light-matter interaction.

 \subsection{Nonlinear response: direct light-matter interaction}    
 We start with calculating corrections generated directly by the 
 time-dependent vector potential. Corrections due to the order 
 parameter oscillations which contain the Higgs mode contribution are considered below in 
 Sec.\ref{SubSec:HiggsModeQuasiclassics}.        
 In the presence of the oscillating external field 
 $\bm A_\Omega e^{i\Omega \tau}$ we can find the solution of 
 Eilenberger equation (\ref{Eq:Eilenberger}) 
 in the form of expansion by the orders of $\bm A_\Omega$ as follows
 %%%%%%%%%%%%%%%%%%% 
 \begin{align} \label{Eq:gAExpansion}
 & \hat g(\tau_1, \tau_2) = 
 \\ \nonumber
 & T\sum_\omega e^{i\omega_1\tau_1} 
 [ \hat g_0(1) e^{-i\omega_1\tau_2}  
 +
 \hat g_{A}(12) e^{-i\omega_2\tau_2}
 +
 \\ \nonumber
 & \hat g_{AA}(13) e^{-i\omega_3\tau_2}
 +
 \hat g_{AAA}(14) e^{-i\omega_4\tau_2}
 ]  
 \end{align}  
 where we introduce the shortened notation to define the 
 frequency dependence such as
 $\hat g_A(ij)= \hat g_A (\omega_i,\omega_j)$ and the shifted Matsubara 
 frequencies $\omega_1 = \omega+2\Omega$, 
  $\omega_2 = \omega+\Omega$, $\omega = \omega$, 
  $\omega_4 = \omega-\Omega$. The zeroth-order solution given by the unperturbed propagator given from Eqs.(\ref{Eq:GFimp},\ref{Eq:QuasiclassicalGF})
   \begin{align} \label{Eq:QCg0}
  \hat g_0(\omega) = \frac{\hat \tau_3 \omega -\hat\tau_2\Delta}{s(\omega)}  
\end{align}     
 where $s(\omega) = \sqrt{\omega^2 + \Delta^2}$.    
 The expansion terms in Eq.(\ref{Eq:gAExpansion}) can be found in the form 
 where momentum direction dependence is explicitly defined  
 \begin{align} \label{Eq:g1aAnsatz}
 & \hat g_A = \alpha \cos\chi \hat g_{1a}(\omega)
 \\ \label{Eq:g2Ansatz}
 & \hat g_{AA} = \alpha^2 (\cos^2\chi - 1/3) 
 \hat g_{2a} + \alpha^2 \hat g_{2s}/3
 \\ \label{Eq:g3Ansatz}
 & \hat g_{AAA} = \alpha^3 \cos\chi\left[ 
 (\cos^2\chi - 1/3) \hat g_{3a} + \hat g_{3s}/3 \right]
 \end{align}
 where we denote $\cos\chi = \bm v_F\bm A_\Omega / v_FA_\Omega$ and 
 $\alpha = e v_F A/c$. 

 The impurity collision integral $[\langle \hat g\rangle, \hat g]$ 
 disappears for the isotropic parts of the propagator, so that we 
 have in the first-order $[\langle \hat g\rangle, \hat g]^{(1)} = 
 [\hat g_0,\hat g_A]$, in the second-order 
 $[\langle \hat g\rangle, \hat g]^{(2)} = 
 \alpha^2 (\cos^2\chi - 1/3)[\hat g_0,\hat g_{2a}]$ and for the third-order correction 
 $[\langle \hat g\rangle, \hat g]^{(3)} = 
 [\hat g_0,\hat g_{3}] + [\hat g_{2s},\hat g_{A}] /3$.
 Then we obtain a chain of equations to determine corrections driven 
 by the direct coupling to the vector potential. 
 The equation for the first-order correction reads
  %%%%%%%%%%%%%%%%%%%   
  \begin{align} \label{Eq:1OrderQcEq}
  & i \left[ \hat\tau_3 \hat g_0(3) - 
  \hat g_0(2) \hat\tau_3\right] =
  \\ \nonumber
  & \tilde s_2\hat g_0(2) 
  \hat g_{1a}(23) - 
  \tilde s_3 \hat g_{1a}(23) 
  \hat g_0(3) ,
  \end{align}
  %%%%%%%%%%%%%%%%%%
  where we introduce the notation $s_i= s(\omega_i)$ and 
  $\tilde s_i = s_i +1/2\tau_{imp}$.
  The second order correction $\hat g_{2s}$ is 
  %%%%%%%%%%%%%%%%%%  
 \begin{align} \label{Eq:2OrderQcEq}
 & i \left[ 
 \hat\tau_3 \hat {g}_{1a}(34) - \hat {g}_{1a}(23)\hat\tau_3  
 \right] = 
 \\ \nonumber
 & s_2 \hat g_0(2) \hat g_{2s}(234) - 
 s_4 \hat g_{2s}(234) \hat g_0(4) .
 \end{align}   
 The equation for $\hat g_{2a}$ has the similar form with the 
 replacements $s_{2,4} \to \tilde s_{2,4} $ in the right 
 hand side. And finally the equation for the third-order correction is 
  \begin{align}  \nonumber
  & \left[ i \hat\tau_3 + \frac{\hat g_{1a}(12)}{2\tau_{imp}}
  \right]
  \hat {g}_{2s}(234) - \hat {g}_{2s}(123) 
  \left[ i \hat\tau_3 + \frac{\hat g_{1a}(34)}{2\tau_{imp}} 
  \right]
   = 
   \\ \label{Eq:3OrderQcEq}
  & \tilde s_1\hat g_0(1) 
  \hat g_{3s} - 
  \tilde s_4 \hat g_{3s} 
  \hat g_0(4) ,
  \\   \label{Eq:3OrderQcEqA}
  & i\left[\hat\tau_3 \hat {g}_{2a}(234) - \hat {g}_{2a}(123)
  \hat\tau_3 \right]
  = 
  \tilde s_1\hat g_0(1)\hat g_{3a} 
  - 
  \tilde s_4 \hat g_{3a}\hat g_0(4) .
  \end{align}
 The solution of above equations reads as 
 follows. The first-order correction is given by 
 \begin{align} \label{Eq:g1a}
 \hat g_{1a} (12) = i \frac{\hat g_0(1)
 \hat\tau_3 \hat g_0(2) - \hat\tau_3}
 {s_1 + s_2 + \tau_{imp}^{-1}} .
 \end{align}   
 The isotropic second-order correction $\hat g_{2s}$ is given by 
 \begin{align} 
 \label{Eq:g2s}
 & \hat g_{2s}(234) = i \frac{s_2 \hat g_0(2) \hat X(234)
 +
 s_4 \hat X(234) \hat g_0(4) }{ s_2^2 - s_4^2 } ,
 \\ \label{Eq:X}
 & \hat X(234) = \hat\tau_3 \hat g_{1a}(34) - \hat g_{1a}(23)\tau_3 
 \end{align}
 %%%  
 and the anisotropic part $\hat g_{2a}$ is obtained by the 
 replacements $s_{2,4} \to \tilde s_{2,4} $ in the right 
 hand side of Eq.(\ref{Eq:g2s}). Finally, the third-order 
 correction is given by 
 \begin{align} \label{Eq:g3s}
 & \hat g_{3s,3a} = i\frac{\tilde s_1 \hat g_0(1)\hat Y_{s,a}
 + \tilde s_4 \hat Y_{s,a} \hat g_0(4)
 }
 {(s_1-s_4)(s_1 + s_4 + \tau_{imp}^{-1})} ,
 \\ \label{Eq:Y3s}
 & \hat Y_s = 
 \\ \nonumber
 & \left[ \hat\tau_3  - \frac{i\hat g_{1a}(12)}{2\tau_{imp}} \right] 
 \hat g_{2s}(234) - \hat g_{2s}(123) \left[ \hat\tau_3 - 
 \frac{i\hat g_{1a}(34)}{2\tau_{imp}} \right] ,
 \\ \label{Eq:Y3a}
 & \hat Y_a = \hat\tau_3 \hat g_{2a}(234) - \hat g_{2a}(123) \hat\tau_3 . 
 \end{align}  
  %%%%%%%%%%%%%%%%%%%%%%%%%%%%%%%%5

 As discussed below in Sec.\ref{Sec:AnalyticalContinuation} in order to use  solutions (\ref{Eq:g3s}) for the numerical calculation it is necessary to convert them into the form which does not have the factor $s_1-s_4$ in the denominators.
 This can be done with the help of normalization condition $\hat g\circ \hat g =1$ which yields the following relations for the corrections 
 \begin{align} \label{Eq:NormalizationG1a}
 & \hat g_0(1) \hat g_{1a} (13) + 
 \hat g_{1a} (13) \hat g_0(3) =0 
 \\ \label{Eq:NormalizationG2s}
 & \hat g_0(1) \hat g_{2s}(123) + \hat g_{2s}(123) 
  \hat g_0(3) + \hat g_{1a}(12)\hat g_{1a}(23) =0
   \\ \label{Eq:NormalizationG3s}
 & \hat g_0(1) \hat g_{3s} + \hat g_{3s} \hat g_0(4) + 
  \\ \nonumber
 & \hat g_{1a}(12)\hat g_{2s}(234) +
  \hat g_{2s}(123)\hat g_{1a}(34) =0
 \end{align}  
 The commutation relations for $\hat g_{2a}$ and $\hat g_{3a}$
 are obtained by substituting these functions instead of 
 $\hat g_{2s}$ and $\hat g_{3s}$ to the relation (\ref{Eq:NormalizationG2s},\ref{Eq:NormalizationG3s}),
 respectively.   As the consistency check, one can prove by the direct calculation that the solutions (\ref{Eq:g2s},\ref{Eq:g3s}) satisfy Eqs.(\ref{Eq:NormalizationG2s},\ref{Eq:NormalizationG3s}). 
 With the help of these relations one can rewrite Eq.(\ref{Eq:g3s})
 in the form suitable for numerics as described in the Appendix \ref{AppSec:s14}. Using this form it is also possible to derive  the diffusive limit consistent with the results given directly by the Usadel equations as discussed in Sec.\ref{SubSec:DirtyLimit} and in the 
 Appendix \ref{App:DirtyLimit}.

  %%%%%%%%%%%%%%%%%%%%%%%%%%%%%%%%%%%%%%%%%%%%%%%%%
 \subsection{Higgs mode contribution to the nonlinear response}
 \label{SubSec:HiggsModeQuasiclassics}
   
 Besides corrections to the GF induced directly by the coupling to vector 
 potential we need to take into account the nonlinear current induced 
 through order parameter oscillations. 
 This process is depicted by the diagrams shown in Fig. 
 \ref{Fig:Paramagnetic}a.      
 First, let us calculate the current generated by the vector potential 
 $\bm A_{\Omega} e^{i\Omega t}$ combined with the order parameter 
 oscillating with the frequency  
 $2\Omega$ which we denote as 
 $\hat \tau_1 \Delta_{2\Omega} e^{i\Omega t}$.
 We can find the corresponding solutions of Eilenberger equation 
 in the form 
 \begin{align}
 & \hat g(\tau_1, \tau_2) = \hat g_{\Delta} (\tau_1,\tau_2) + 
 \hat g_{\Delta A}(\tau_1, \tau_2) 
 \\
 & \hat g_{\Delta} (\tau_1,\tau_2)=  T \sum_\omega \hat g_{\Delta} (13)
 e^{i\omega_1\tau_1 - i \omega_3\tau_2}
 \\
 & \hat g_{\Delta A} (\tau_1,\tau_2)=  \alpha\cos\chi T\sum_\omega
 \hat g_{\Delta A} (1234) 
 e^{i\omega_1\tau_1 - i \omega_4\tau_2} .
 \end{align}    
  
 We use a similar approach to solve the chain of equations for corrections 
 as in the previous section. Then the obtained first- and second-order 
 solutions read    
 \begin{align} \label{Eq:g1Delta}
 &\hat g_{\Delta} (13) = \Delta_{2\Omega}
 \frac{ \hat g_0(1)\hat\tau_2 \hat g_0(3)-\hat\tau_2}
 {s_1 + s_3 }
 \\    \label{Eq:gADelta}
 & \hat g_{A\Delta} = i \frac{
 \tilde s_1 \hat g_0(1)
 (\hat Y_\Delta-i\hat Y_A)
 +
 \tilde s_4 
 (\hat Y_\Delta-i\hat Y_A) \hat g_0(4)
 }
 {(s_1-s_4)(s_1 + s_4 + \tau_{imp}^{-1})} 
 \\ 
 & \hat Y_A=\Delta_{2\Omega}[\hat\tau_2 \hat g_{1a}(34) - 
  \hat g_{1a}(12)\hat\tau_2 ]
 \\ \label{Eq:YDelta}
 & \hat Y_\Delta = 
 \\ \nonumber
 & \left[ \hat\tau_3  - \frac{i\hat g_{1a}(12)}
 {2\tau_{imp}} \right] \hat g_{\Delta}(24) 
 - \hat g_{\Delta}(13) \left[ \hat\tau_3  - \frac{i\hat g_{1a}(34)}
 {2\tau_{imp}} \right] ,
 \end{align}  
 where the first-order correction due to the vector potential 
 $\hat g_{1a}$ is given by the Eq.(\ref{Eq:g1a}). 
 Note that the correction $\hat g_\Delta$ induced by the order parameter 
 oscillations is completely isotropic and therefore is not affected by the 
 impurity scattering collision integral. From this one can immediately conclude that the polarization operator which 
 determines the Higgs mode is not sensitive to the disorder. 
 
 As discussed below in Sec.\ref{Sec:AnalyticalContinuation} in order to use the solution (\ref{Eq:gADelta}) for the numerical calculation of the analytical continuation it is necessary to convert it into the form which does not have the factor $s_1-s_4$ in the denominator.
 This can be done with the help of normalization condition of the quasiclassical GF. For the corrections it yields 
 \begin{align} \label{Eq:NormalizationGDelta}
 & \hat g_0(1) \hat g_\Delta (13) + 
 \hat g_\Delta (13) \hat g_0(3) =0 
 \\ \label{Eq:NormalizationGADelta}
 & \hat g_0(1) \hat g_{A\Delta} + \hat g_{A\Delta} \hat g_0(4) +
 \\ \nonumber
 & \hat g_{1a}(12)\hat g_\Delta (24) + 
 \hat g_\Delta (13)\hat g_{1a}(34)   =0
 \end{align}  
 One can check by the direct calculation with the certain analytical effort that the solutions (\ref{Eq:g1Delta},\ref{Eq:gADelta}) satisfy the conditions (\ref{Eq:NormalizationGDelta},\ref{Eq:NormalizationGADelta}). 
 Then, using these relations it is possible to convert (\ref{Eq:NormalizationGADelta}) 
 to the required form as described in the Appendix \ref{AppSec:s14}. 
  
 The correction $\hat g_{A\Delta}$ is the basic building block to calculate  
 the Higgs mode-related current $j_H$ according to the
 diagram in Fig.\ref{Fig:Paramagnetic}b.
 In order to obtain $j_H$ as the third-order response to the external field 
 we need to calculate the order parameter amplitude $\Delta_{2\Omega}$
 excited in the second order by 
 $\bm A_\Omega$ as shown by the diagram in 
 Fig.\ref{Fig:Higgs}a. This can be done in two steps described below.

 %%%%%%%%%%%%%%%%%%%%%%%%%%%%%%%%%%%%%%
 {\bf External perturbation of the order parameter.}
 First, from the diagram in Fig.\ref{Fig:Higgs}a we find the source,
 that is the time-dependent order parameter induced directly by the 
 external field. 
 The amplitude $F_\Delta$ can be found using the isotropic 
 second-order correction for the GF 
 $\hat g_{2s}$  calculated above (\ref{Eq:g2s})
 \begin{align} \label{Eq:FDelta0}
 F_\Delta (2\Omega) =  
 - \frac{\lambda\alpha^2}{3} T \sum_\omega 
 {\rm Tr} [ \hat\tau_2 \hat g_{2s} ] .
 \end{align}
 %      

 %%%%%%%%%%%%%%%%%%%%%%%%%%%%%%%%%%%%%%%%%%%%%%%%%%%%%%       
 {\bf Polarization operator}
 Second, to find the order parameter amplitude $\Delta_{2\Omega}$ driven 
 by the external source $F_\Delta$ we need to take into account the 
 polarization corrections given by the diagrammatic series shown in Fig.
 (\ref{Fig:Polarization}). Thus the renormalized order parameter vertex  
 $\hat{{\cal D}}$ denoted in Fig.(\ref{Fig:Polarization}) by the 
 filled triangle is related to the bare one $\hat S_0$ as 
 $\hat{ {\cal D}} = \hat{ {\cal D}}_0 /[1-\Pi(2\Omega)]$. 
 Here $\Pi$ is the polarization operator given by the single bubble 
 in Fig. \ref{Fig:Polarization}. 
 Thus the total order parameter perturbation is related to the 
 external source $F_\Delta$ as 
 \begin{align} \label{Eq:Delta2OmF}
 \Delta_{2\Omega} = \frac{F_\Delta}{1-\Pi(2\Omega)} .
 \end{align}     
 Previously, the polarization operator has been calculated in the clean  
 limit \cite{PhysRevB.92.064508, Cea2016}. 
 In the presence of impurities the diagram summation becomes more 
 complicated because besides the modification of propagator lines 
 we need also to take into account the impurity ladders. However, as 
 shown below in result the expression for $\Pi$ remains the same as in 
 the clean limit.  
  
 Instead of the direct diagram summation it is much faster to 
 calculate the  polarization operator in the presence of disorder within the
 quasiclassical formalism. Let us assume that there is an external driving 
 term in the gap function given by $e^{2i\Omega \tau} F_\Delta \hat\tau_1$. 
 Then in the Eilenberger equation we have the source 
 $ i\hat\tau_3\hat\tau_1 e^{2i\Omega\tau} F_\Delta = -\hat\tau_2 i e^{2i\Omega\tau} F_\Delta $. 
 Besides that there is a response of the order parameter which we denote  
 $i e^{2i\Omega\tau}  \Delta_{2\Omega} \hat\tau_1$. Then the amplitude of 
 the total off-diagonal driving term in the Eilenberger equation is 
 $\Delta_{2\Omega} + F_\Delta$. Then we can use this amplitude instead of 
 $\Delta_{2\Omega}$ in the expression (\ref{Eq:g1Delta}) for the 
 corresponding correction to the propagator. 
 The self-consistency equation $ \Delta_{2\Omega} = - \lambda T \sum_
 \omega {\rm Tr} \hat\tau_2 \hat g_{\Delta}(13)$ yields
 %%%
 \begin{align} \label{Eq:PolariazationOp0}
 \Pi (\Omega) = 1+ \lambda T \sum_\omega 
 \left( \frac{s_1s_3 -\Delta^2 +\omega_1\omega_3}{s_1s_3 (s_1+s_3)} -
 \frac{1}{s_1} \right)
 \end{align}
 where we denote as  before $s_1 = \sqrt{\omega_1^2 +\Delta^2}$ and 
 $s_3 = \sqrt{\omega_3^2 +\Delta^2}$. Taking into account that 
 $\sum_\omega ( s_3^{-1} - s_3^{-1}) =0$ the expression for polarization 
 operator (\ref{Eq:PolariazationOp0}) can be transformed to 
 %
 %%%%%%%%%%%%%%%%%%%%%       
 \begin{align}
 \Pi(2\Omega) =  
 1+  \lambda T \sum_\omega \frac{\Delta^2 + \Omega^2}
 {s(\omega^2 - \Omega^2)}
 \end{align} 
 The analytical continuation of this expression to real 
 frequencies as explained in Sec.(\ref{Sec:AnalyticalContinuation}) 
 yields the result coinciding with the one obtained previously in 
 the clean limit 
 \cite{Volkov1973,Kulik1981,PhysRevB.26.4883,PhysRevB.92.064508,Cea2016}.  
 Note however, that here we have demonstrated that this expression is valid 
 for arbitrary impurity scattering time. 
    
 \subsection{Total current}
 Finally we collect expressions for the current 
 $j = j_{AAA} + j_H$   
 given by the diagrams shown in 
 Fig.\ref{Fig:Paramagnetic}
 \begin{align} \label{Eq:jAAAFin}
 & j_{AAA} (\Omega) = - i \pi\nu e v_F \frac{\alpha^3}{9} 
 T\sum_\omega {\rm Tr} [\hat\tau_3
( \hat g_{3s} +  4 \hat g_{3a}/5 )]  ,
 \\ \label{Eq:jHFin}
 & j_H (\Omega) =  - i \pi\nu e v_F \frac{\alpha}{3} T\sum_\omega 
 {\rm Tr} [\hat\tau_3\hat g_{\Delta A} ] .
 \end{align}
 In order to find the current at real frequency we need to make analytical 
 continuation of Eqs.(\ref{Eq:jAAAFin},\ref{Eq:jHFin})
 as described in Sec.\ref{Sec:AnalyticalContinuation}. 
 In general, this procedure leads to the expressions which can be handled  only numerically and in Sec.\ref{Sec:Numerics} we show the characteristic dependencies of THG current on various parameters. 
 However, in a number of limiting cases it is possible to treat these expressions analytically which we discuss in Sec.\ref{Sec:Estimations}.      
 In the next Sec. we prove that the results of quasiclassical calculations of the non-linear response coincide with those obtained by the summation of diagrams with current vertices in the presence of impurity self-energy and
 ladders.  

 %%%%%%%%%%%%%%%%%%%%%%%%%%%%%%%%%%%%%%
   \subsection{Analytical continuation} 
 \label{Sec:AnalyticalContinuation}
 %%%%%%%%%%%%%%%%%%%%%%%%%%%%%%%%% 
 In order to find the real-frequency response we need to implement 
 the analytical continuation of Eq.(\ref{Eq:jAAAFin},\ref{Eq:jHFin}). 
 These third-order responses are obtained by the summation of expressions 
 which depend on the four shifted fermionic frequencies such as 
 $g = g(\omega_1,\omega_2,\omega_3,\omega_4)$. 
 The analytical continuation of the sum by Matsubara frequencies 
 is determined according to the general rule \cite{kopnin2001theory}
 %%%%%%%%%%%%%%%%
  \begin{align} \label{Eq:AnalyticalContinuationGen}
  & T\sum_\omega g(\omega_1,\omega_2,\omega_3,\omega_4)
  \to  
  \\ \nonumber
  & \sum_{l=1}^4 \int \frac{d\varepsilon}{4\pi i}  
  n_0(\varepsilon_l)
  \left[ g(...,  -i \varepsilon^R_l, ...) - 
  g(...,  -i \varepsilon^A_l, ...)  \right]
  \end{align}
 %%%%%%%%%%%%%%%%
 where $n_0(\varepsilon) = \tanh (\varepsilon/2T ) $ is the equilibrium  
 distribution function. In the r.h.s. of (\ref{Eq:AnalyticalContinuationGen}) we substitute in each term 
 $\omega_{k<l} = - i\varepsilon^R_k$ and 
 $\omega_{k>l} = - i\varepsilon^A_k$ for $k=1,2,3,4$,
  denote 
  $\varepsilon_k = \varepsilon + (3-k)\Omega$ 
  and  $\varepsilon^R = \varepsilon + i\Gamma$, $\varepsilon^A = 
  \varepsilon -i \Gamma$. Here the term with $\Gamma>0$ is added to   
  shift of the integration  contour into the corresponding half-plane.
  At the same time,  $\Gamma$ can be used as the Dynes 
  parameter \cite{dynes84}
  to describe the effect of different depairing mechanisms 
  on spectral functions in the superconductor.
  We implement the analytical continuation in such a way that 
  %%%%   
  $
  s(-i\varepsilon^{R,A}) 
  = - i \sqrt{ (\varepsilon^{R,A})^2- \Delta^2}
   $    assuming that the branch cuts run from $(\Delta,\infty)$
   and $(-\infty, -\Delta)$. 
   
   A special care should given to the differences $s_i-s_j$
   in the denominators of Eq.(\ref{Eq:g2s},\ref{Eq:g3s},\ref{Eq:gADelta}) 
   for the second- and third-order responses.
   When analytically continued to the real energies and frequencies these combinations become zero for certain energies. Indeed, e.g. for $i=1$, $j=4$ we have $\varepsilon -2\Omega = - (\varepsilon +\Omega)$ is $\varepsilon = \Omega/2$ so that $s_1^A = s_4^R$ for such energy.
   Thus, the numerical integration of the expressions that contain 
   combinations like $s_1^A-s_4^R$ in the denominators is not possible.
   Fortunately, the expressions (\ref{Eq:g2s},\ref{Eq:g3s},\ref{Eq:gADelta})  with certain analytical effort can be written in the form which does not contain $s_i-s_j$ combinations in the denominators. 
   The procedure  of how this can be done using the commutation relations (\ref{Eq:NormalizationG1a},\ref{Eq:NormalizationG2s},
\ref{Eq:NormalizationG3s}) and (\ref{Eq:NormalizationGDelta},
\ref{Eq:NormalizationGADelta}) for the second-order corrections 
$\hat g_{2s,2a}$ and third-order correction $\hat g_{3s,3a}$, $\hat g_{A\Delta}$ is given in the Appendix \ref{AppSec:s14}.   
 %%%%%%%%%%%%%%%%%%%%%%%%%%%%%%%%%%%%%%%%%                       

  %%%%%%%%%%%%%%%%%%%%%%%%%%%%%%%%%%%%%%%%%          
        
  \section{Diagram summation}
  \label{Sec:DiagramSummation}
  It is the goal of this section to demonstrate that corrections  
  given by the diagrams with current 
  vertices can be calculated using quasiclassical approximation 
  introduced in the previous section. For this purpose we consider 
  impurity ladder diagrams for corrections up to the third order in external field and use them to derive 
  equations for the corresponding momentum-integrated propagators 
  introduced according to Eq.(\ref{Eq:QuasiclassicalGF}). 
  
  \subsection{First-order corrections}
  \label{SubSec:1OrderCorr}
\begin{figure}[htb!]
 \centerline{$
 \begin{array}{c}
  \includegraphics[width=0.9\linewidth]{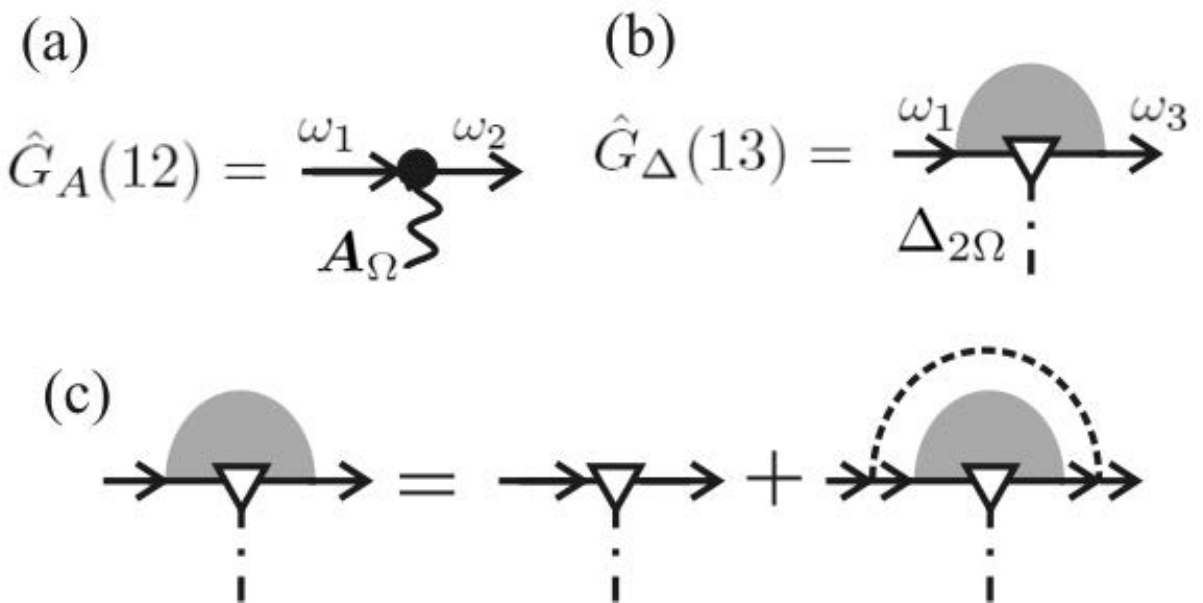}
  % \put (-310,15) {\LARGE $\hat G^{(2)}_{AA}=$} %{\LARGE (c) }
  \end{array}$}
 \caption{\label{Fig:GF1order} (Color online) 
  Diagrammatic representation of the first order corrections 
  due to (a) interaction  with external field $\bm A_{\Omega}$ through the current vertex and (b) interaction with the order parameter perturbation 
$\Delta_{2\Omega}$  through the order parameter vertex dressed by the impurity ladder shown by the dashed region. 
  (c) Diagrammatic equation for the correction (b).  }
 \end{figure}

  Let us start with the simplest diagrams for the first-order 
  corrections induced by the interaction with 
  external field $\bm A_\Omega e^{i\Omega t}$ through the current-type vertex  and by the order 
  parameter modulation  $\Delta_{2\Omega} e^{2i\Omega t}$ as 
  shown in Fig.(\ref{Fig:GF1order}). 
  
    %%%%%%%%%%%%%%%%%%%%%%%%%%%%%%%%%%%%%%%%%%%%%%%
 %%%%%%%%%%%%%%%%%%%%%%%%%%%%%%%%%%%%%%%%%%%%%%%%%%%  
  
 We aim to demonstrate the general approach for deriving equation for 
 the momentum-integrated propagators using the simplest diagram 
 shown in Fig.\ref{Fig:GF1order}a.
 Let us introduce the notation 
 %%%%%%%%%%%%%%%%%%%%%%%%%%%
 \begin{align} 
 \hat g_{A}=\frac{i}{\pi}
 \int d\xi_p \hat\tau_3  \hat G_{A} .
 \end{align}
 %%%%%%%%%%%%%%%%%%%%%%%%%%%
 The key idea of the derivation of the simplified equation for 
 $\hat g_{A}(12)$ is to use the following trick. Let us multiply 
 the function $\hat G_A (12)$ by $\hat G_0^{-1} (1)$ from the left
 and by $\hat G_0^{-1} (2)$ from the right, subtract the results 
 and integrate by $\xi_p$. 
 We use that  Eq.(\ref{Eq:GFimp}) yields the relations 
 $ \hat G_0^{-1} (j) = \tilde{\Delta}_j \hat\tau_1 
 + i \tilde{\omega}_j \hat\tau_0 + \xi_p \hat\tau_3 $   
 and 
 $\tilde{\Delta}_j \hat\tau_1 + i \tilde{\omega}_j \hat\tau_0 = 
 i (s_j + 1/2\tau_{imp})\hat g_0(\omega_j)\hat\tau_3$,
 where $s_j = \sqrt{\omega_j^2 + \Delta^2} $.
 Then we eliminate off-shell contributions in the momentum integrals 
 to express the result through quasiclassical propagators 
 \begin{align} \label{Eq:lhs}
 & \int \frac{d\xi_p}{\pi} \left[ \hat G_0^{-1} (1) 
 \hat G_{A}(12) - \hat\tau_3 \hat G_{A} (12) 
 \hat G_0^{-1} (2) \hat\tau_3 \right]         
 = 
 \\ \nonumber
 & \tilde s_1\hat g_0(1) \hat g_A - 
 \tilde s_2\hat g_A \hat g_0(2) 
 \end{align}
 where we introduce the notation $\tilde s_j = s_j + 1/2\tau_{imp}$. 
The obtained expression coincides with the r.h.s. 
of the Eilenberger Eq.(\ref{Eq:1OrderQcEq}) for the correction $\hat g_{A}$. 
Next let us derive the l.h.s. of the equation for $\hat g_A$. 
Using the diagram Fig.\ref{Fig:GF1order}a we get that 
$\hat G_0^{-1} (1) \hat G_{A}(12) - \hat\tau_3 \hat G_{A}(12) \hat G_0^{-1} (2)  \hat\tau_3= \alpha \cos\chi [ \hat G_0(2) - \hat\tau_3\hat G_0(1)\hat\tau_3] $. Then, integrating by $\xi_p$ we obtain the equation to determine $\hat g_A (12)$ coinciding with Eqs.(\ref{Eq:g1aAnsatz}, \ref{Eq:1OrderQcEq}).
          
  %%%%%%%%%%%%%%%%%%%%%%%%%%%%%%%%555
  Let us now consider the momentum integrated correction 
  \begin{align} 
  \hat g_{\Delta}=\frac{i}{\pi}
  \int d\xi_p \hat\tau_3  \hat G_{\Delta} 
  \end{align}
  where $\hat G_{\Delta} $ is given by the more complicated 
  diagram Fig.\ref{Fig:GF1order}b with ladder insertion. 
  Following the procedure described above we get the 
  r.h.s. of the equation for $\hat g_{\Delta}(13)$ in the form 
  similar to (\ref{Eq:lhs}). To obtain the l.h.s one needs to 
  use the diagrammatic equation Fig.(\ref{Fig:GF1order}c).
  Using the fact that $\hat G_{\Delta}$ does not depend on the 
  momentum direction we write this equation in the algebraic 
  form as follows
  %%%%%%%%%%%%%%%%%%%%%%%% 
 \begin{align} \label{Eq:IntegralGDelta}
 & \hat G_{\Delta}(13) = \hat G_{0}(1)\hat\Delta_{2\Omega}
 \hat G_{0}(3) + 
 \hat G_{0}(1) 
 \frac{ \hat g_\Delta \hat\tau_3}
 {2i\tau_{imp}}
 \hat G_{0}(3)
 \end{align}     
 %%%%%%%%%%%%%%%%%%%%%%%
 %        
 From these equations we obtain
 $\hat G_0^{-1} (1) 
 \hat G_{\Delta}(13) - \hat\tau_3 \hat G_{A}(13) 
 \hat G_0^{-1} (3)\hat\tau_3= 
 (\hat \Delta_{2\Omega} -i \hat g_\Delta\hat
 \tau_3/2\tau_{imp} ) \hat G_0(3) - 
 \hat\tau_3\hat G_0(1)( \hat \Delta_{2\Omega}\hat\tau_3 -
 i \hat g_\Delta/2\tau_{imp} )  $.
 Then terms with impurity scattering in l.h.s. and r.h.s. appear to be exactly  
 the same. In result we recover the quasiclassical equation for 
 $\hat g_\Delta$ with the solution Eq.(\ref{Eq:g1Delta}) which is 
 not sensitive to impurity scattering.   
   
 \subsection{Second-order correction }
 \label{Eq:NonLinearVertex}
\begin{figure}[htb!]
 \centerline{$
 \begin{array}{c}
 \includegraphics[width=0.9\linewidth]{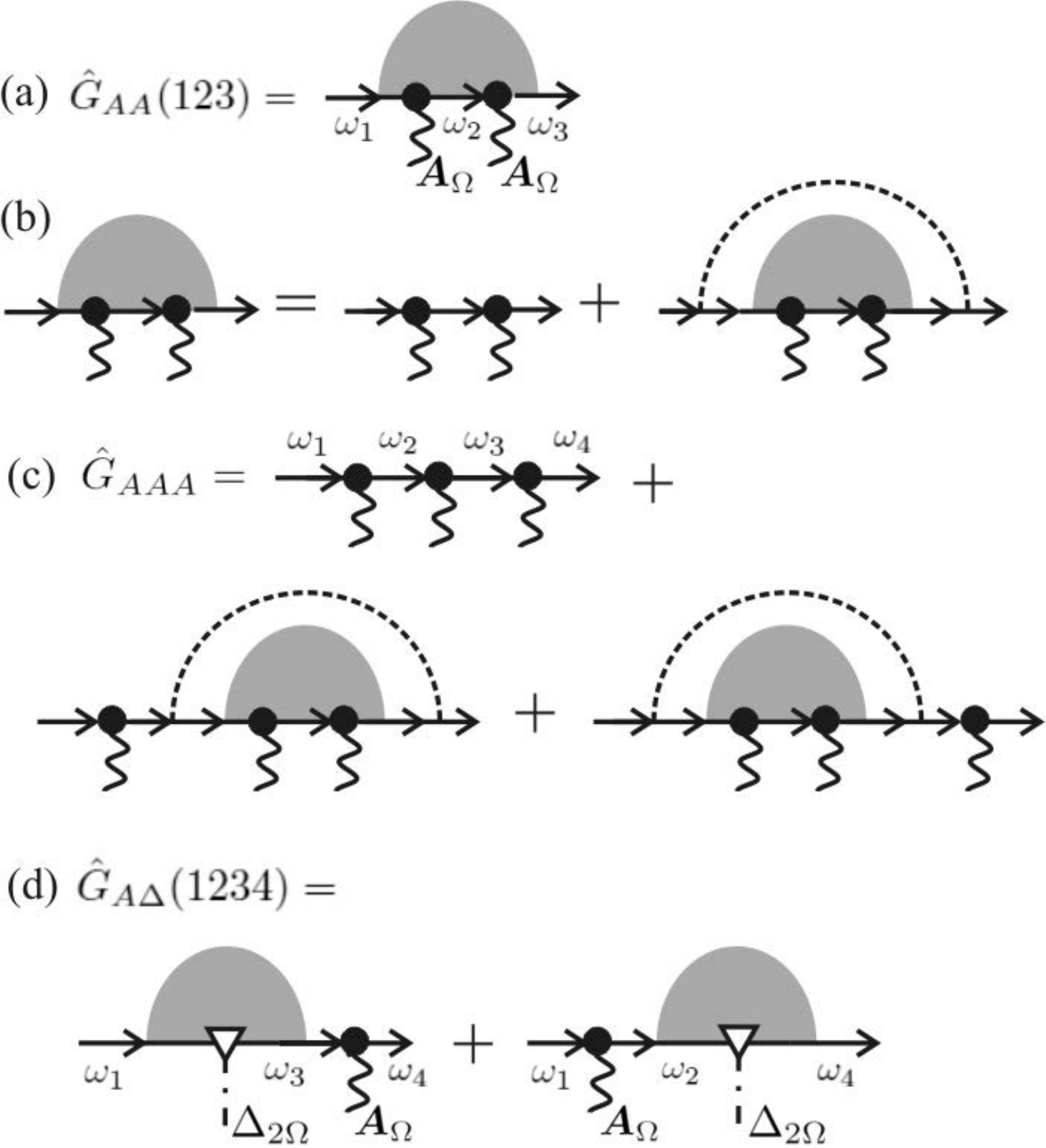}
 \end{array}$}
 \caption{\label{Fig:GF2order} 
 (a) The second-order correction $G_{AA}$ due to the direct coupling 
 to vector potential $\bm A_\Omega$  through
current vertices, $\omega_k=\omega +(3-k)\Omega$. 
 (b) Diagrammatic equation to determine the second-order correction for the propagator with current vertices 
modified by impurity scattering in ladder approximation. (c) The third-order correction $G_{AAA}$. (d) Correction that is generated by the combined order parameter and vector potential perturbations. 
 }
 \end{figure}

 In this section we calculate the basic element of the
 nonlinear response diagrams Fig.\ref{Fig:Paramagnetic}, that 
 is the second-order correction $\hat G_{AA}$ modified by the impurity 
 ladder as shown in Fig.\ref{Fig:GF2order}a. 
 We denote the frequencies $\omega_1 = \omega+2\Omega$, $\omega_2 
  = \omega+\Omega$, $\omega_3 = \omega$, 
  $\omega_4 = \omega-\Omega$.

 %%%%%%%%%%%%%%%%%%%%%%%%%%%%%%%%%%%%%%%%%%%%%%%
 %%%%%%%%%%%%%%%%%%%%%%%%%%%%%%%%%%%%%%%%%%%%%%%%%%%  
              
   The correction $\hat G_{AA}$ can be though of as a result of 
 the second-order electron-photon interaction process. It is determined by the equation 
 shown diagrammatically in Fig.\ref{Fig:GF2order}b. 
 Due to the presence of two current vertices  the 
 angle average associated with the dashed impurity line 
 produces non-zero result.              
   The corresponding integral equation for $\hat G_{AA}$ reads
 %
 %%%%%%%%%%%%%%%%%%%%%%%% 
 \begin{align} \label{Eq:IntegralG2}
 & \hat G_{AA}(123) = \alpha^2\cos^2\chi \hat G_{0}(1)
 \hat G_{0}(2)\hat G_{0}(3)
 + 
 \\ \nonumber
 &  \hat G_{0}(1) 
 \frac{\langle \hat g_{AA}(123)\rangle_\chi}{2i\tau_{imp}} 
 \hat\tau_3 G_{0}(3)
 \end{align}     
 %%%%%%%%%%%%%%%%%%%%%%%
 where we denote the angular average 
 $\langle f \rangle_\chi = \int_0^\pi d\chi \sin\chi f(\chi) $ and the 
 momentum-integrated GF $\hat g_{AA} = i\pi^{-1} \int d\xi_p \hat G_ {AA}$.
 From the Eq.(\ref{Eq:IntegralG2}) we can obtain equation for the 
 momentum-integrated GF $\hat g_{AA}$. Using the same method as described
 above in Sec.\ref{SubSec:1OrderCorr} we get the equation for 
 $\hat g_{AA}=\hat g_{AA}(123)$ which reads as
 %%%%%%%%%%%%%%%%%%%%%%%%%%%%%%%  
 \begin{align}
 & i\alpha^2 \cos^2\chi [ \hat g_{1a}(12) \hat\tau_3 - 
 \hat\tau_3 \hat g_{1a}(23)] =
 \\ \nonumber
 & \tilde s_1 \hat g_0(1)\hat g_{AA} - \tilde s_3 \hat g_{AA}\hat g_0(3) 
 +
  \frac{\langle \hat g_{AA} \rangle_\chi\hat g_0(3) - \hat g_0(1)\langle  
 \hat g_{AA} \rangle_\chi}{2\tau_{imp}}
 \end{align}
 %%%%%%%%%%%%%%%%%%%%%%%%%%%%%% 
 Using the ansatz (\ref{Eq:g2Ansatz}) we get from here Eq.(\ref{Eq:g2s}) for the components $\hat g_{2s}$ and $\hat g_{2a\\
 }$. 
  
 %%%%%%%%%%%%%%%%%%%%%%%%%%%%%%%%%%%%%%%%%%% 
 \subsection{Third-order correction}
 The third-order correction to the GF is determined by the diagram shown in 
 Fig. \ref{Fig:GF2order}c . The corresponding analytical expression reads 
 \begin{align} \label{Eq:3orderIntegral}
 & \hat G_{AAA} = - \alpha^3\cos^3\chi 
 \hat G_0(1)\hat G_0(2)\hat G_0(3)\hat G_0(4) - 
 \\ \nonumber
 & \frac{\alpha\cos\chi}{2i\tau_{imp}}
 [ \hat G_0(1) \langle \hat g_{AA}(123) \rangle_\chi \hat\tau_3 \hat G_0(3)\hat G_0(4) +
 \\ \nonumber 
 & \hat G_0(1)\hat G_0(2) \langle \hat g_{AA}(123) \rangle_\chi
 \hat\tau_3 \hat G_0(4) ]
 \end{align}
     
 The straightforward calculation of $\hat G_{AAA}$ is rather lengthy. 
 However, it is possible to show that the momentum-integrated function
 $\hat g_{AAA} = i\pi^{-1} \int d\xi_p \hat G_ {AAA}$
 satisfies Eilenberger Eqs. (\ref{Eq:3OrderQcEq},\ref{Eq:3OrderQcEqA}).
 The key idea of this derivation is to
use the same trick as introduced above in Sec.\ref{SubSec:1OrderCorr} to rewrite the diagrammatic equation in terms of the quasiclassical
propagators. First, we multiply Eq.(\ref{Eq:3orderIntegral}) from the left by $\hat G_0^{-1}(1)$, from the right by $\hat G_0^{-1}(4)$ and then subtract the
results in the same way as given by the Eq.(\ref{Eq:lhs}) to obtain the r.h.s. of the Eilenberger equation in the form
\begin{align} \label{Eq:lhs3}
 & \int \frac{d\xi_p}{\pi} \left[ \hat G_0^{-1} (1) 
 \hat G_{AAA}- \hat\tau_3 \hat G_{AAA} 
 \hat G_0^{-1} (4) \hat\tau_3 \right]         
 = 
 \\ \nonumber
 & \tilde s_1\hat g_0(1) \hat g_{AAA} - 
 \tilde s_4\hat g_{AAA} \hat g_0(4) 
 \end{align}
The l.h.s. of the resulting equation can be expressed in terms of the quasiclassical propagators  directly from Eq.(\ref{Eq:3orderIntegral}).
In this way we obtain
  \begin{align} \label{Eq:Qcd3order}
    & i\alpha \cos\chi [ \hat Y_a/3 + (\cos^2\chi -1/3)\hat Y_s ] = 
    \\ \nonumber
    &  \tilde s_1\hat g_0(1) \hat g_{AAA} - 
 \tilde s_4\hat g_{AAA} \hat g_0(4) 
  \end{align}
 where $\hat Y_{s,a}$ are given by the Eqs.(\ref{Eq:Y3s},\ref{Eq:Y3a}). 
 Using the ansatz (\ref{Eq:g3Ansatz}) we get from Eq.(\ref{Eq:Qcd3order}) 
 the solutions for components (\ref{Eq:g3s}).

 %%%%%%%%%%%%%%%%%%%%%%%%%%%%%%%%%%%%%%%%%%%%%% 
 \subsection{Corrections due to the Higgs mode }    
   
Correction to the GF $\hat G_{A\Delta}$  generated by the combined action of the vector potential and the time-dependent order
parameter is given by the diagram in Fig.\ref{Fig:GF2order}d. This perturbation determine the non-linear current $j_H$ which is sensitive
to the excitation of the Higgs mode as shown by the diagram in Fig.2b. As before we are interested in the momentum-integrated
function $\hat g_{A\Delta}=i\pi^{-1}
 \int d\xi_p \hat\tau_3  \hat G_{A\Delta}$ because it determines the correction to the current. Treating the diagram
Fig.\ref{Fig:GF2order}d using exactly the same procedure as above we arrive to the equation for $\hat g_{A\Delta}$ obtained directly from the
Eilenbeger formalism. Its solution is given by the Eq.(\ref{Eq:gADelta}).

  %%%%%%%%%%%%%%%%%%%%%%%%%%%%%%%%%%%%%%%%    

 \section{Limiting cases and estimations}
 \label{Sec:Estimations}
 \subsection{Normal state or very high frequencies $\Omega \gg \Delta$} 
 First, let us check that in the normal state the  third-harmonic response 
 disappears. This can be seen from the analytical continuations of the Eq. 
 (\ref{Eq:jAAAFin}) for $j_{AAA}$ because the Higgs mode-related part $j_H$ 
 does not exist in the normal state. Let us introduce the condensed 
 notation e.g. $g(-i \varepsilon_{1}, -i \varepsilon_2, -i \varepsilon_3, -
 i \varepsilon_4) = g (1^R,2^R,3^R,4^R )$.
 In the normal state $\hat g_0(-i\varepsilon^{R,A} ) = \pm \hat\tau_3 $. 
 Besides that we have $s (-i\varepsilon_k^R) = - i [\varepsilon +  (k-3) 
 \Omega] $ and $s (-i\varepsilon_k^A) =  i [\varepsilon +  (k-3) \Omega] $ 
 so that the sum $s (-i\varepsilon_k^R) + s (-i\varepsilon_n^A) =  i (n-k)
 \Omega$, that is does not depend on energy. 
  
 Then for the third-order corrections given by Eqs.(\ref{Eq:g3s}) one can 
 see that $\hat g_{3a}(1^R,2^R,3^R,4^R) =  \hat g_{3s}(1^A,2^A,3^A,4^A)=0$. 
 Moreover, $\hat g_{3a} (1^A,2^A,3^R,4^R) = - 
 2 \hat g_{3a} (1^A,2^R,3^R,4^R) = - 2 \hat g_{3a} (1^A,2^A,3^A,4^R)$. The 
 same relations are true for $\hat g_{3s}$. Due to this relations, the 
 contributions from different branch cuts 
 (\ref{Eq:AnalyticalContinuationGen}) in the expression for the normal 
 state current $j_{AAA}$ cancel identically. 
 
 Note that for the frequencies $\Omega \gg \Delta$ the effect of superconducting correlations disappears, so one can consider the system as normal metal. Hence the non-linear response vanishes in this high-frequency limit. This is the reason why diagrams with current vertices can be neglected e.g. when studying Raman response where the frequencies associated with single external lines are assumed to be rather large. 
 
  %%%%%%%%%%%%%%%%%%%%%%%%%%%%%%%%%%%%%%%%%%%%%% 
 \subsection{Absence of impurities $\tau_{imp}=\infty$}    
 \label{SubSec:NoImpurity}
 Previously it has been noted that in the clean limit, in the spatially 
 homogeneous case and finite frequency the order parameter amplitude is 
 not affected by the irradiation \cite{Gorkov1968}. In that work by 
 Gor'kov and Eliashber only the contribution of diagrams with current 
 vertices has been taken into account. Under the same assumptions 
 non-linear current response also disappears. 
 Here we check our general results for consistency against this limiting 
 case of $\tau_{imp}=\infty$.  
     
 {\bf Order parameter perturbation.}
 First, let us look at the second-order corrections which determine 
 perturbation of the order parameter $F_\Delta$ according to the diagram 
 Fig.\ref{Fig:Higgs}a. In the limit $\tau_{imp}=\infty$ the solutions 
 isotropic second-order correction to GF  (\ref{Eq:g2s}) yields (see Appendix \ref{SecApp:NoDisorderNoHiggs} for details)
 \begin{align} \label{Eq:Trg2Clean}
 & {\rm Tr} [\hat\tau_2 \hat g_{2s} (123) ]= 
 -\frac{\Delta}{2\Omega^2} \left( 2s_2^{-1} - s_1^{-1} -  
 s_3^{-1}\right)
 \end{align}
   Therefore in the Eq.(\ref{Eq:FDelta0})  the sum over frequencies disappears so that
  \begin{align} \label{Eq:F1fin}
 F_\Delta (2\Omega) = 
 \lambda \alpha^2 \frac{ \Delta}{2\Omega^2} T\sum_\omega 
 (2s_2^{-1} - s_1^{-1} - s_3^{-1})    = 0.
 \end{align}

 Now let us consider the contribution of the diagram with density modulation
 Fig.\ref{Fig:Higgs}b. In this case we obtain 
 %%%%% 
 \begin{align} \label{Eq:FDeltaDnParabolic}
 F^{(d)}_\Delta = 
 \tilde{\lambda}\frac{e^2 
 A^2 }{2mc^2} {\rm tr}
 [ \hat\tau_1\hat G_0(1) \hat\tau_3 \hat G_0(3)  ]
 \end{align}
 %%%%%
 where ${\rm tr}[\hat X] = 
 T\sum_\omega \int d^3\bm p/(2\pi)^3 {\rm Tr} [\hat X]$. 
 The momentum integral in (\ref{Eq:FDeltaDnParabolic}) disappears
 due to the particle-hole symmetry\cite{PhysRevB.92.064508,Cea2016} which 
 holds up to the corrections of the order $T_c/E_F \ll 1$. 
 The same conclusion holds in the presence of impurity scattering which modifies propagators and the density vertex as shown in Fig.\ref{Fig:Higgs}b by the shaded region. 

Thus, the contribution of density modulation to the Higgs mode excitation
always vanishes due to the particle-hole symmetry. Therefore it has been claimed that the Higgs mode does not contribute to the third harmonic generation \cite{Cea2016,Cea2018}.
Below we show that in the presence of impurities the non-zero $F_{\Delta}$ 
is obtained within the quasiclassical approximation and does not contain 
any such small prefactors . 
    
 {\bf THG current response.}
 Let us now consider expression for the current $j_{AAA}$  which in the 
 absence of impurities is determined by the 
 third-order correction $\hat g_{AAA}$ given by Eqs.(\ref{Eq:g3Ansatz}). 
 The expression which determined the current can be written as 
 \begin{align} \label{Eq:g3nImp} 
 {\rm Tr} [\hat\tau_3 \hat g_{3}] = 
 \frac{i}{6\Omega^2} [f(123) -f(234)]
 \end{align}     
 where $f(123) = f(\omega_1,\omega_2,\omega_3)$ is the 
 function which exact form is rather lengthy and not particularly important. 
 Then due to the Eq.(\ref{Eq:g3nImp}) the sum over Matsubara frequencies in   
 the Eq.(\ref{Eq:jAAAFin}) for current disappears since 
 $\sum_\omega f(123) = \sum_\omega f(234)$. 
 
 %%%%%%%%%%%%%%%%%%%%%%%%%%%%%%%%%%%%%%%
 \subsection{Transition to the clean limit $\tau_{imp}T_c\gg 1$.}   
 As shown above in Sec.\ref{SubSec:NoImpurity} the finite-frequency contribution of diagrams with current vertices Fig.\ref{Fig:Higgs}a
disappears in the absence of impurity scattering and other relaxation mechanisms. The contribution of  diagram with density vertices Fig.\ref{Fig:Higgs}b is zero with the accuracy of the particle-hole symmetry
near the Fermi level.
So that the contribution of Higgs mode excitation to the THG signal is expected to have the negligible amplitude\cite{Cea2016,Cea2018} as compared to the direct coupling described by the diagram in 
Fig.\ref{Fig:Diamagnetic}.
 Thus the polarization dependence of THG signal is expected to be sensitive to the lattice anisotropy.
Here we show that the above conclusions can drastically altered in the presence of rather small impurity concentrations,
typical for all realistic superconducting samples, especially in thin film geometries. Our goal is to find the threshold value of impurity scattering when the diagrams with current vertices become dominant and so that the
quasiclassical theory described in Sec.\ref{Sec:QuasiclassicalFormalism} is applicable for calculating non-linear  properties.

 To understand the magnitude of this threshold in this subsection let us analyse the amplitudes in the regime of 
 low frequencies, impurity scattering rates and small temperatures as compared to the critical 
 temperature 
 $T,\Omega,  \tau^{-1}_{imp} \ll T_c$. In Sec.\ref{Sec:Numerics}
 the numerical results with broad range frequency and scattering rate dependencies are presented.  

 %%%%%%%%%%%%%%%%%%%%%%%%%%%%%%%%%%%
 {\bf Order parameter perturbation.}
First, let us estimate the magnitude of the external order parameter perturbation in the presence of weak disorder. 
 The contribution of diagrams with current vertices Fig.\ref{Fig:Higgs}a 
 to the order parameter perturbation $F_\Delta$ can be obtained using 
 solution (\ref{Eq:g2s}) expanded in the regime 
 $\Omega,  \tau^{-1}_{imp} \ll T_c$
 %%%%%%%%%%%%%%%%%%%%%%% 
 \begin{align}
 {\rm Tr} [\hat\tau_2\hat g_{2s}] = \frac{\Delta}{6}
 \left[ \frac{\Delta^2 - 2\omega^2}
 {(\Delta^2 + \omega^2)^{5/2} } + \frac{\omega^2 -\Delta^2}{\tau_{imp} 
 (\Delta^2+ \omega^2)^3}  \right]
 \end{align}
 %%%%%%%%%%%%%%%%%%%%%%
 Then at small temperatures $T \ll T_c$ the first term here vanishes upon the integration over $\omega$
 while the second term yields the leading order expansion by the 
 scattering rate 
 \begin{align} \label{Eq:FDeltaExpansion}
 F_\Delta = \frac{\lambda}{144} \frac{\alpha^2}{\Delta\tau_{imp}}
 \end{align}
 Thus the Higgs mode amplitude perturbation driven by the linear electron-photon coupling terms in the Hamiltonian is non-zero for the finite concentration of impurities.

 %%%%%%%%%%%%%%%%%%%%%%%%%%%%%%%%%%%%%%%%%%%%%%  
 { \bf THG current response.}
 At non-zero frequencies the current $j_{AAA}$ given by diagrams with current vertices Fig.(\ref{Fig:Paramagnetic}) disappears without 
impurity scattering and other relaxation mechanisms. 
At the same time the  current $j_{AAA}^{(d)}$ determined by 
diagrams with density vertices in  Fig.\ref{Fig:Diamagnetic} remains 
non-zero. 
Let us find the threshold value of impurity scattering when the latter contribution $j_{AAA}^{(d)}$ can be neglected.
 
It is convenient to introduce dimensionless amplitudes $I_{AAA}$, $I_{H}$ 
of the currents (\ref{Eq:jAAAFin},\ref{Eq:jHFin}) determined by the contribution of diagrams with current vertices 
\begin{align} \label{Eq:jAAAHdimensionless}
 & I_{H, AAA}= - \frac{j_{H, AAA}}{j_0} 
 \\ \label{Eq:j0}
 & j_0 = \frac{e\nu}{T_c^2} \left( \frac{v_F e A}{c} \right)^3 ,
  \end{align}
  Here $j_0$ is the normalization current density. 
  In the same way, the contribution of the density modulation diagram in Fig.\ref{Fig:Diamagnetic} can be written as in terms of the dimensionless amplitude 
 \begin{align}\label{Eq:IdAAA}
 & I^{(d)}_{AAA}= - \frac{j^{(d)}_{AAA}}{j^{(d)}_0} 
 \\ \label{Eq:jd0}
  & j^{(d)}_{0}  = 
 em\nu  \left(\frac{e A}{mc} \right)^3  .
 \end{align}
 Then the dimensionless amplitude is given by 
 \begin{align} \label{Eq:jAAADensDimensionless}
  I^{(d)}_{AAA} (\Omega) = 
 \frac{\Delta^2}{8\Omega} 
 \pi T\sum_\omega \frac{s_1 - s_3}{s_1s_3(\omega_1 + 
 \omega_3)}  
\end{align}  
 %
 % \begin{align}\label{Eq:jAAADensDimensionless}
% & j^{(d)}_{AAA} (\Omega) = 
% em\nu  \left(\frac{e A}{mc} \right)^3  I^{(d)}_{AAA}  
% (\Omega)
% \\
% & I^{(d)}_{AAA} (\Omega) = 
% \frac{\Delta^2}{8\Omega} 
% \pi T\sum_\omega \frac{s_1 - s_3}{s_1s_3(\omega_1 + 
% \omega_3)}  
% \end{align}
 %%%
% where we denote  $\omega_1 = \omega + 2\Omega$, 
% $\omega_3 =\omega$, 
% $s_{1,3} = \sqrt{\omega_{1,3}^2 + \Delta^2}$. 
 %   
 Hence from the Eqs.(\ref{Eq:jAAADensDimensionless},\ref{Eq:jAAAHdimensionless}) the ratio of different contributions to the current can be expressed through the ratio of dimensionless amplitudes as follows
 \begin{align} \label{Eq:RatioLowT}
 \frac{j^{(d)}_{AAA}}{j_{H,AAA}} = 
 \left(\frac{T_c}{2E_F}\right)^2 
 \frac{I^{(d)}_{AAA}}{I_{H,AAA}} ,
 \end{align}
 where the prefactor is determined by $j_0/j_0^{(d)} =T_c^2/2E_F^2 $ and  $E_F=mv_F^2/2$ is the Fermi energy.  
 As we mentioned above, in the limit $\tau_{imp}\to \infty$ 
 the amplitude $I_{AAA}$ disappears while 
 $I^{(d)}_{AAA}$ remains finite. However, due to the very 
 small  prefactor in Eq.(\ref{Eq:RatioLowT})
 $(T_c/E_F)^2 \gg 1$ the threshold level of $\tau_{imp}$
 occurs to be quite large. 
 
 To understand the magnitude of this threshold let us analyse the amplitudes in the regime of 
 low frequencies, impurity scattering rates and small temperatures as compared to the critical 
 temperature 
 $T,\Omega,  \tau^{-1}_{imp} \ll T_c$.

 Under the above assumption one can obtain the analytical 
 expression for 
 the density modulation-induced amplitude  $I^{(d)}_{AAA}$. 
 For small $\Omega$ we can replace 
 $\Omega^{-1} (s_3^{-1} - s_1^{-1}) = -2 d s^{-1}_3 /d\omega = 
 \omega (\omega^2 + \Delta^2)^{-3/2}$. So that the current 
 amplitude becomes just $ I^{(d)}_{AAA} = 1/2$. 
  
 The calculation of quasiclassical contribution is more involved. 
 Let us implement the Taylor expansion in terms of the frequency
 $\Omega$ and the scattering rate $\tau^{-1}_{imp}$ of 
 the Eqs.(\ref{Eq:g2s},\ref{Eq:g3s}). In this way we obtain the leading-order terms 
 \begin{align} 
 \label{Eq:g3aExpansion}
 & \hat g_{3a} = -  \frac{i}{2}
 \frac{\Delta^4 - 4 \Delta^2 \omega^2}
 {(\Delta^2 + \omega^2)^{7/2}} 
 +
 \frac{3 i}{4\tau_{imp}} 
 \frac{\Delta^4 - 4 \Delta^2 \omega^2}
 {(\Delta^2 + \omega^2)^4} 
 \\ 
 \label{Eq:g3sExpansion}
 & \hat g_{3s} = -  \frac{i}{2}
 \frac{\Delta^4 - 4 \Delta^2 \omega^2}
 {(\Delta^2 + \omega^2)^{7/2}} 
 +
 \frac{i}{\tau_{imp}} 
 \frac{\Delta^4 - 2 \Delta^2 \omega^2}
 {(\Delta^2 + \omega^2)^4} 
 \end{align}
 
 Since we consider low temperatures, when calculating contributions 
 to the current the frequency summation has to be replaced by the 
 integral, e.g. $2\pi T\sum_\omega \hat g_{3a} = \int \hat g_{3a} d\omega$. Then  the first terms in the r.h.s. of expansions 
 (\ref{Eq:g3aExpansion},\ref{Eq:g3sExpansion})
 disappears so we end up with the result 
 %%%%%%%%%%%%%%%5 
 \begin{align}
 -i  T \sum_\omega \hat g_{3s} = 
 -4 i T \sum_\omega \hat g_{3a} =  \frac{3}{32\Delta^3 }
 \end{align}
 %%%%%%%%%%%5%%%
 Thus we get the leading-order contribution to the quasiclassical current amplitude 
 %%%%%%%%%%%%%%%%%% 
 \begin{align} \label{Eq:IqAAA}
 I_{AAA} \approx \frac{10^{-3}}{\tau_{imp}T_c}
 \end{align}
 %%%%%%%%%%%%%%%%%%
 where we took into account the relation $\Delta = 1.76 T_c$.
 Thus from (\ref{Eq:RatioLowT}) we get the ratio of the density-
 modulation and quasiclassical currents given by 
 \begin{align} \label{Eq:RatioLowTResult}
 \frac{j^{(d)}_{AAA}}{j_{AAA}} \approx 10^3 (\tau_{imp} T_c)  
 \left(\frac{T_c}{E_F}\right)^2 
 \end{align}
   Note that Eqs.(\ref{Eq:IqAAA},\ref{Eq:RatioLowTResult}) are valid at small frequencies $\Omega \ll \Delta$ and they also agree with the static limit when $\Omega =0$. Indeed, in the static limit $j_{AAA}+j_H$ determines non-linear correction to the Meisser current. This correction can be shown to vanish at $T=0$ in the absence of disorder $\tau_{imp}=\infty$ in agreement with Eq.. 
     
 Based on the estimation (\ref{Eq:RatioLowTResult}) the contribution of  diagram with density vertex becomes 
 dominating in the limit determined by  the condition 
 %%%
 \begin{align} \label{Eq:ConditionDiaParaLowT}
 \tau_{imp}T_c > 10 ^{-3} \left(\frac{E_F}{T_c}\right)^2
 \end{align}      
 %%% 
 Taking into account that in usual superconductors like NbN with 
 $T_c\approx 10 K$ and $E_F \approx 10^4 K$ the above condition 
 yields $ \tau_{imp}T_c > 10 ^{3} $. 
 This criterion means that the superconductor should be in the 
 super-clean regime
 \cite{PhysRevLett.79.1377, PhysRevB.51.15291} defined as $\tau_{imp}T_c > E_F/T_c 
 \approx 10^3 $. Up to now the only known system where this regime is realized\cite{PhysRevB.51.15291,PhysRevB.44.9667} is the superfluid He$^3$ which generically does not contain any impurities.
 In solid state systems the certain amount of disorder is always present. 
 Besides that in thin films the scattering time is bounded from 
 above by the time of flight of electrons between interfaces. 
 Taking into account that in the clean limit the coherence length 
 is $\xi = T_c/v_F$ the above criterion means that the 
 film should be thicker than $100 \xi$ which is much larger than 
 what has been used in experiments \cite{Matsunaga1145,
 PhysRevLett.120.117001,1809.10335}. 
 Besides that, typical materials used in THz spectroscopy  
 experiments like NbN superconductors usually have strong intrinsic disorder so the dirty limit $\tau_{imp}T_c <1$ is realized there even without taking into account the interface scattering. 
  
  {
    Another interesting tendency characteristic for the transition to the clean limit $\tau_{imp}T_c \gg 1$ is that  the Higgs mode-related current 
  is suppressed much strongly than the other component so that $j_H\ll j_{AAA}$. 
 This can be understood from the 
 expansion b small parameter $(\tau_{imp}T_c)^{-1}\ll 1$ of the GF correction 
 $\hat g_{A\Delta}$ which determine $j_H$ according to the 
 Eq.(\ref{Eq:jHFin}).  In this case the amplitude of 
 the order parameter external perturbation is given by 
 (\ref{Eq:FDeltaExpansion}), that is it already contains the small 
 parameter $(\tau_{imp}T_c)^{-1} \ll 1$.
 Besides that expanding $\hat g_{A\Delta}$ given by 
 Eqs.(\ref{Eq:g1Delta},\ref{Eq:gADelta},\ref{Eq:YDelta}) we get
 %%%%%%%%%%%%% 
 \begin{align} \label{Eq:gADeltaExpansion}
 \hat g_{A\Delta} = i \Delta \Delta_{2\Omega} 
 \left[ 
 \frac{ ( \Delta^2 - 2 \omega^2)}{(\Delta^2 + \omega^2)^{5/2}}  
 -
 \frac{\Delta^2- \omega^2}{\tau_{imp}(\Delta^2 + \omega^2)^3 }  
 \right] .
 \end{align}
 %%%%%%%%%%%%
 The first term here vanishes as usual upon the integration by 
 $\omega$ while the second term together with 
 Eqs.(\ref{Eq:FDeltaExpansion},\ref{Eq:Delta2OmF})
 yields the amplitude of the current $j_H$. Here we need to take 
 into account the low-frequency asymptotic of the polarization operator 
 which follows from Eq.(\ref{Eq:PolariazationOp0}) 
 $\Pi (\Omega\ll \Delta) =1 - \lambda  $. The we get the non-linear current 
 generated due to the Higgs mode  excitation with the  dimensionless 
 amplitude given by 
 %%%%%%%%%%%%%%%%%%%% 
 \begin{align}
 |I_H| \approx  2\;\cdot 10^{-3} \left(\frac{T_c}{\Delta}\right)^4 \frac{1}{(\tau_{imp}T_c)^2} ,
 \end{align}
 %%%%%%%%%%%%%%%%%%% 
where at low temperatures $(T_c/\Delta)^4 \approx 0.1$.  
 Thus one can see that the suppression of $I_H (\tau_{imp})$ in the  
 transition to clean case is determined by the second order of the small parameter $(\tau_{imp}T_c)^{-1} \ll 1$. Therefore in this limit 
 $I_H \ll I_{AAA}$ except of the vicinity if the Higgs mode resonance at $\Omega = \Delta$ where the amplitude $I_H$ is enhanced by the factor $\sqrt{\Delta/\Gamma}$, where $\Gamma$ is the Dynes parameter. However, if the impurity scattering is sufficiently weak $(\tau_{imp}T_c)^{-1} < \sqrt{\Gamma/\Delta} $
 the direct contribution to nonlinear current $I_{AAA}$ dominates for all frequencies.   These different regimes are illustrated by the numerical results below in the section\ref{Sec:Numerics}. 
   }

 Estimations that we provided above rule out the necessity to consider the 
 contribution of density-vertex diagrams to describe the non-linear 
 properties of known superconducting materials. Besides that the thin-film 
 samples used for the non-linear response studies in the THz regime are 
 generically in the dirty regime, because the mean free path is bounded 
 from above by thickness because the surface scattering of electrons 
 mimics impurity scattering. 
 Therefore  in realistic superconducting  samples 
 where the impurity scattering rate 
 is always above the superclean limit the non-linear response is determined only by  the diagrams with current vertices 
 Fig.\ref{Fig:Paramagnetic}. As shown above this means that one can use the quasiclassical approximation with impurity collision integrals which yields major simplification as compared to the direct summation of diagrammatic series.  Besides that usually the low-temperature superconductors are in the dirty regime 
 $\tau_{imp}T_c \ll 1$ which can be treated within even simpler  Usadel theory as discussed in the next section. However,  the general solutions we obtain can be applied with some modifications to calculate nonlinear responses in clean materials like the d-wave high temperature superconductors \cite{PhysRevLett.120.117001} or iron-based superconductors 
 which are in the regime $\tau_{imp}T_c >1$.

 %%%%%%%%%%%%%%%%%%%%%%%%%%%%%%%%%%%%%%%%%%%%%
 \subsection{Dirty limit $\tau_{imp} T_c \ll 1$}
 \label{SubSec:DirtyLimit}
 Previously,  non-linear electromagnetic properties of 
 superconductors have been studied mostly in the diffusive system 
 using the Usadel formalism 
 \cite{Gorkov1968,Gorkov1969,GorkovTDGL1968,Eliashberg1970,
 Ivlev1971,PhysRevLett.118.047001,PhysRevB.97.184516}. 
 The general case with arbitrary impurity concentration has been 
 discussed in the  linear  response regime\cite{Artemenko:1979}. 
 Here we present for the first time calculations of the non-linear 
 responses with arbitrary impurity  scattering time. Therefore it is 
 important to establish connection with the Usadel theory results 
 which should be obtained from our general expressions in the limit 
 $\tau_{imp}T_c \ll 1$. 
 That is, the isotropic second-order correction $\hat g_{2s}$ which 
 determines the order parameter perturbation and the expression for 
 non-linear current response can be obtained directly from 
 Eqs.(\ref{Eq:UsadelGen},\ref{Eq:MWCI},\ref{Eq:CurrentUsadel})
 as $(\alpha^2 D/v_F^2)g_{2}(123) $ where
 %%%%%%%%%%%%%%%%%%%%%%  
 \begin{align} 
   \label{Eq:g2Usadel}
 & g_{2}(123) =  \frac{\hat\tau_3\hat g_0 (2)\hat\tau_3 - \hat g_0 (1)\hat\tau_3 \hat g_0 (2) \hat\tau_3 \hat g_0(3)}{s_{1} + s_{3}}
  \\ 
 \label{Eq:Current3OmegaUsadel}
 & \frac{j_{AAA}}{j_{3\Omega}} =  
  T \sum_\omega {\rm Tr}
 \hat\tau_3 [ \hat g_0(1) \hat\tau_3 \hat g_{2}(234) + 
  \hat g_0(4)\hat\tau_3 \hat g_{2}(123)  ]
 \\ \label{Eq:CurrentH3OmegaUsadel}
  & \frac{j_H}{j_{3\Omega}} =
   T \sum_\omega {\rm Tr}
 \hat\tau_3 [ \hat g_0(1) \hat\tau_3 \hat g_{\Delta}(24) + 
 \hat g_0(4)\hat\tau_3 \hat g_{\Delta}(13)  ] .
 \end{align} 
  %%%%%%%%%%%%%%%%%%%%%%%
    %
    where we normalize the current by the amplitude
    $j_{3\Omega} = \pi e^2 (A_\Omega/c)^3 D\sigma $. 
    The correction $\hat g_\Delta$ induced by the order parameter oscillation is given by Eq.(\ref{Eq:g1Delta}) and does not depend on the impurity scattering rate. 
 In the Appendix (\ref{App:DirtyLimit}) we demonstrate that 
 the same expressions follow from the general 
 Eqs.(\ref{Eq:g2s},\ref{Eq:jAAAFin}) as the leading term expansions 
 by the small parameter $\tau_{imp} T_{c} \ll 1$.

  %%%%%%%%%%%%%%%%%%%%%%%%%%%%%%%%%%%%%%%%%%%%%%%%%%%%%%%%       
 \section{Numerical results and discussion}
  \label{Sec:Numerics}

%%%%%%%%%%%%%%%%%%%%%%%%%%%%%%%%%%%%%%%%%%%%%%%%
 \begin{figure*}[htb!]
 \centerline{$
 \begin{array}{c}
 \includegraphics[width=1.0\linewidth]{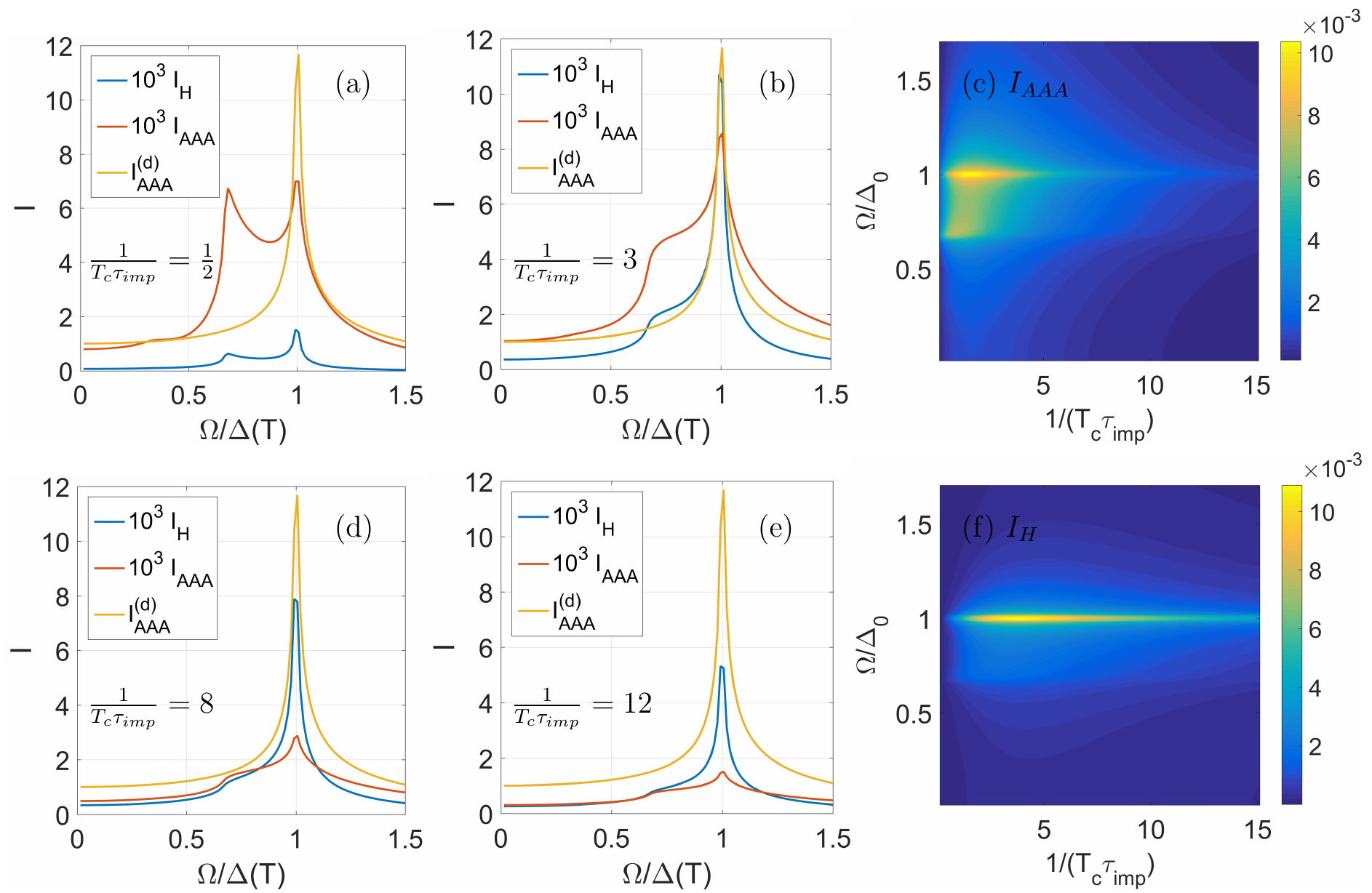}  
 \end{array}$}
 \caption{\label{Fig:JOmScGamma001}
  (a,b,d,e) Absolute values of the THG current amplitudes $I_{AAA}$, 
   $I_{H}$ and $I_{AAA}^{(d)}$
 as functions of $\Omega$ for different values of the impurity scattering rates $(\tau_{imp}T_c)^{-1}$ varying between the clean and dirty regimes. (c,f) Dependencies of $I_{AAA}$ and $I_H$
 on the external frequency and scattering rate. 
 The Dynes parameter $\Gamma=0.01$ and $T=0.1 T_c$ for all plots. 
   }
 \end{figure*}   
 %%%%%%%%%%%%%%%%%%%%%%%%%%%%%%%%%%%%%%%%%%%%%%%%%%%  

  { Analytical estimations in previous section  are obtained at small frequencies $\Omega \ll \Delta$. To understand the full frequency dependencies we implement the numerical calculation using the analytical continuation procedure explained in Sec. \ref{Sec:AnalyticalContinuation}. 
At first, our aim is to compare the dimensionless amplitudes of the three contributions 
        to the current $I_{AAA}$, $I_H$ and $I_{AAA}^{(d)}$
for different values of the parameters. 
        The results are presented in Fig.\ref{Fig:JOmScGamma001} for  different values of the scattering rate varying between the clean and dirty regimes. Here we consider the regime of small temperatures $T=0.1T_c$ and the Dynes parameter is $\Gamma=0.01$. These plots confirm qualitative conclusions made above. In the clean case $(\tau_{imp}T_c)^{-1}=0.5$ (Fig.\ref{Fig:JOmScGamma001}a) the Higgs contribution is much smaller than the direct coupling one $I_H\ll I_{AAA}$. This is despite the fact that Higgs mode contribution is resonant and its maximal value scales like $\sqrt{\Delta/\Gamma}$. The $I_{AAA}$ contribution also has peaks
        both as $\Omega =\Delta$ and $\Omega =2\Delta/3$ although their amplitude does not diverge with $\Gamma \to 0$. The origin of these peaks is the BCS density of states singularity at the energy $\varepsilon =\pm \Delta$. The external radiation at  frequencies  $\Omega =\Delta, 2 \Delta/3$ causes transition between these states with the enhanced probability due to the large density of states. 
        
        With increased scattering rate the general amplitude of the Higgs mode contribution rises so that its maximal value becomes much larger than the direct contribution. For the considered value of Dynes parameter $\Gamma= 0.01$ this happens at rather large scattering $(\tau_{imp}T_c)^{-1}\approx 10$ as shown in Figs.\ref{Fig:JOmScGamma001}d,e. At such parameters $I_H\approx 10^{-3} I_{AAA}^{(d)}$ which means that $j_H\gg j_{AAA}^{(d)}$ if one recalls the overal factor $(T_c/E_F)^2 \approx 10^{-6}$ in front of the density modulation current (\ref{Eq:RatioLowT}). The non-resonant peak in $I_{AAA}$ at $\Omega=\Delta $ remains although the one at $\Omega=2\Delta/3 $ is eliminated by the impurity scattering. The overall dependencies of $I_{AAAA}$
  and $I_H$ on frequency and scattering rates as shown in \ref{Fig:JOmScGamma001}c,f. One can see that first they increase with $(\tau_{imp}T_c)^{-1}$ but then start to decrease.
  We discuss the regime with large scattering rate below using the diffusive limit approach .    
        }  
  
    %%%%%%%%%%%%%%%%%%%%%%%%%%%%%%%%%%%%%%%%%%%%%%%%
 \begin{figure}[htb!]
 \centerline{$
 \begin{array}{c}
 \includegraphics[width=1.0\linewidth]{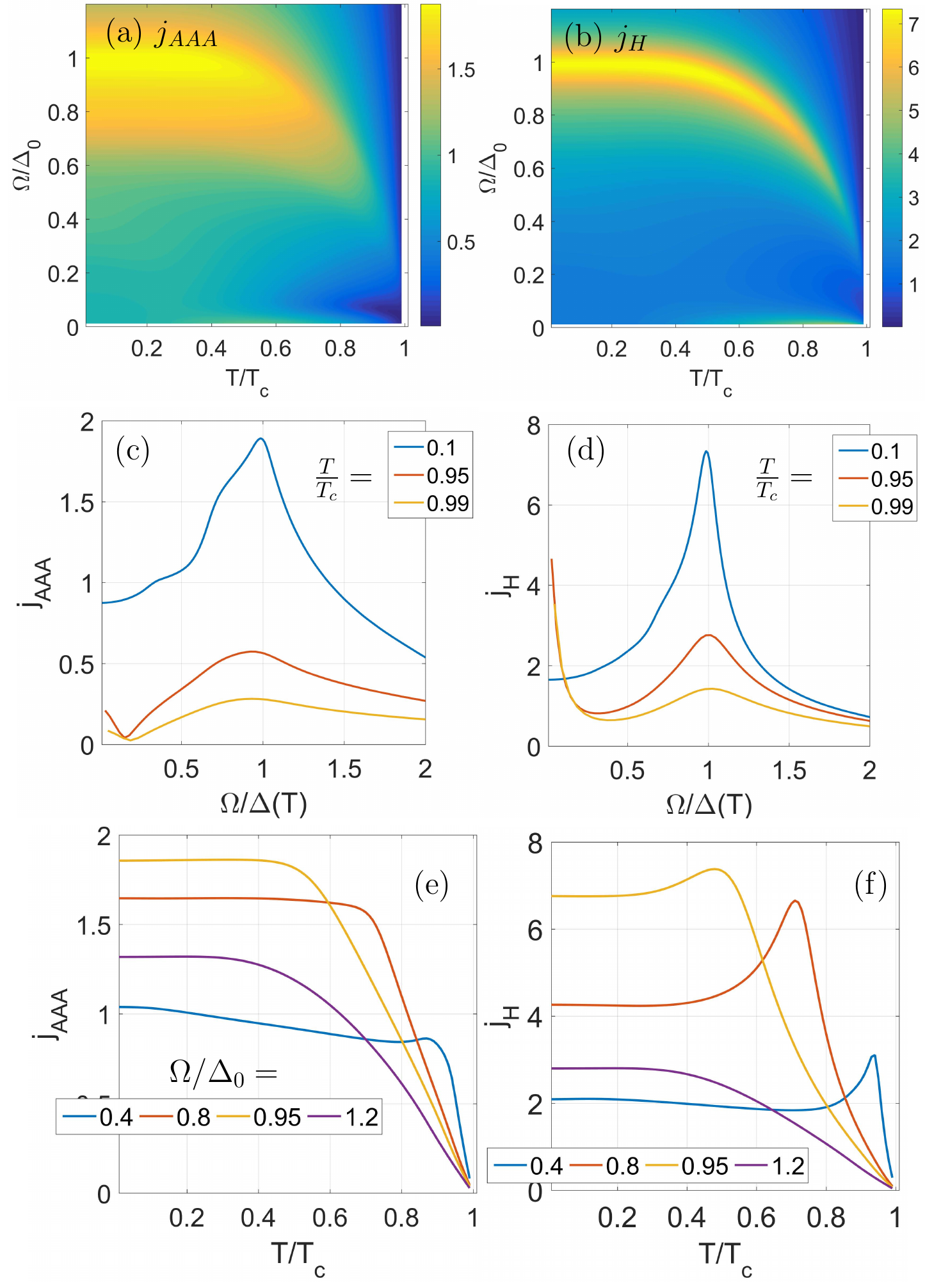}  
 \end{array}$}
 \caption{\label{Fig:Gamma01}
   Absolute values of the THG current components $j_{AAA}$, $j_{H}$ 
 as functions of $T,\Omega$, the Dynes parameter $\Gamma=0.1$. Left panels (a,c,e): the contribution 
 of $j_{AAA}$ current.  Right panels (b,d,f): the contribution 
 of $j_{H}$ current. In (c,d) the curves correspond 
 to different values of temperature. 
 In (e,f) the curves correspond to the different frequencies. 
 The current are normalized by the amplitude $j_{3\Omega}$, see Eqs.(\ref{Eq:Current3OmegaUsadel},\ref{Eq:CurrentH3OmegaUsadel}).
  }
 \end{figure}   
 %%%%%%%%%%%%%%%%%%%%%%%%%%%%%%%%%%%%%%%%%%%%%%%%%%% 

  In thin-film NbN samples used for the non-linear response measurement the amount of scattering is typically rather large. Therefore we consider the diffusive limit below using the Usadel theory results from Sec.\ref{SubSec:DirtyLimit} to calculate non-linear currents. Besides that since NbN has strong electron-phonon coupling \cite{PhysRevB.96.020505} resulting in the 
  enhanced inelastic relaxtion we use larger value of Dynes parameter $\Gamma=0.1$ as compared to the previous example.   
 The results for temperature and frequency dependence of currents   $j_{AAA}$ and $j_H$ are shown 
 in Fig.\ref{Fig:Gamma01}.
   As one can see in Fig.\ref{Fig:Gamma01}a,b   both currents $j_{AAA}$ and $j_H$ have peaks at $\Omega =\Delta (T)$. 
 However, the peak value of Higgs contribution is several times larger. It occurs when the denominator in Eq.(\ref{Eq:Delta2OmF}) reaches 
 it minimal value of 
 $1-\Pi(2\Delta) \approx \sqrt{\Gamma/\Delta}$. 
 So the Higgs-mode related part of non-linear current has the same maximal 
 amplitude $j_{H}(\Omega =\Delta)\propto j_{3\Omega}\sqrt{\Delta/\Gamma}$ as compared  the 
 $j_{AAA}(\Omega =\Delta)\propto j_{3\Omega}$ part. 
 These estimation agrees with the numerical result in 
 Fig.\ref{Fig:Gamma01}c,d.
   
 For higher values of the Dynes parameter which can mimic the enhanced inelastic relaxation in superconductors with strong electron-phonon interaction like NbN \cite{PhysRevB.96.020505} the resonant Higgs peak broadens and decreases. This tendency is illustrated in Fig.\ref{Fig:Gamma01}. The broadened peaks featured by the temperature dependencies of $j_H(T)$ at fixed frequencies (Fig.\ref{Fig:Gamma01}f) 
 are similar to those obtained in the experiment \cite{Matsunaga1145}. At the same time the dependencies of $j_{AAA}(T)$in Fig.\ref{Fig:Gamma01}e show no peaks at all.
 
As we discussed above, the presence of impurities triggers the non-linear response and Higgs mode generation. However, as one can see from the sequence of the plots in Fig.\ref{Fig:JOmScGamma001}c,f  above certain threshold
scattering rate  the amplitudes $I_{H}$ and $I_{AAA}$ start to  
 decrease with decreasing $\tau_{imp}$. This agrees with the  
 diffusive limit results Eqs.(\ref{Eq:Current3OmegaUsadel},
 \ref{Eq:CurrentH3OmegaUsadel}) where the currents are determined 
 by the amplitude has an additional small parameter 
 $(\tau_{imp}T_c)^2$ as compared
 to the prefactor in (\ref{Eq:jd0}) so that 
 $j_{3\Omega} \propto e\nu (eA/c)^3 (\tau_{imp} E_F)^2$.  
 At the same time the density modulation-related current $j_{AAA}^{(d)}$ is not sensitive for disorder and therefore should dominate in the very dirty system as well as in the very clean one. 
 Thus comparing $j_{3\Omega}$ with the prefactor in Eq.(\ref{Eq:jd0}) we obtain that in the diffusive limit $j_{AAA}^{(d)}$
 is negligible as long as $\tau_{imp} E_F>>1$ and start to dominate in the opposite case, that is close to the localization threshold.

 The technique developed in the present work for the arbitrary scattering rates can be applied as well to study Higgs mode generation in non-trivial superconductors  like those with the d-wave symmetry of the order parameter\cite{PhysRevLett.120.117001}. 
 Within quasiclassical formalism such states are described with the help of the anisotropic pairing constant $\lambda_d (\theta,\theta^\prime) = \lambda \sin(2\theta)\sin(2\theta^\prime)$. Here $\theta$ and $\theta^\prime$ are the angles corresponding to the momenta of interacting electrons. Correspondingly the order parameter acquires momentum dependence $\Delta (\theta) \propto \sin(2\theta)$ which should be taking into account when solving the Eilenbeger equation in d-wave superconductor.
  With that one can see that in the absence of impurities the same limitations on the non-linear response  pertain as for the s-wave superconductor. That is, in the completely pure system the second-order electron-photon coupling through the linear terms $V_1$ in the Hamiltonian (\ref{Eq:V1}) does not excite Higgs mode and does not produce any THG signal. 
 The presence of impurities certainly helps the situation although their effect is a bit more tricky than in the isotropic s-wave considered here. However since d-wave pairing can be found only in the clean regime $(T_c\tau_{imp})^{-1} \ll 1$ one can expect that the amplitude of Higgs mode contribution should be strongly suppressed according to the Fig.\ref{Fig:JOmScGamma001}a. However,
 to figure out the resulting amplitude it is necessary to figure out the value of Dynes parameter which can be much smaller than in low-temperature superconductors thus allowing for large $I_H$ peaks 
 even in the clean system. At the same time we don't expect the direct coupling current $j_{AAA}$ to feature pronounced peaks because of the lack of the density of states singularities in the d-wave superconductor.   
 Thus the influence of impurities on the collective modes in  
 superconductors with non-trivial pairing is potentially very 
 interesting although its detailed study  is  beyond the 
 scope of the present paper.
  
 Another interesting  direction which can be addressed using the 
 formalism developed by us is the nonlinear response and
 generation of collective modes in multiband superconductors like 
 MgB$_2$ and iron-pnictide compounds. 
 Here in addition to the impurity scattering important effects can 
 be related to the interband tunnelling of quasiparticles and 
 Cooper pairs which should significantly affect non-linear 
 response. With that we can address recent experimental results on the THz pump-probe experiments with MgB$_2$  \cite{Giorgianni2019}.   
 
 %%%%%%%%%%%%%%%%%%%%%%%%%%%%%%%%%%%%%%%%%%%%%%%%%%%%%%% 
 
 \section{Conclusions}
  We have studied non-linear electromagnet response of superconductors with  the amount of disorder varying between the completely pure limit and  the dirty regime. 
  The impact of our study is threefold. 
  First, it it demonstrated that the quasiclassical approximation allows for the correct description of the non-linear effects in superconductors coupled to the external electromagnetic field. 
  Propagators obtained by solving quasiclassical Eilenberger equation with the impurity collision integral coincide with those obtained by the direct  summation of diagrams with current vertices taking into account the impurity self-energies and ladders.
  
  Second, we demonstrated that effective Higgs mode excitation 
  is possible in usual BCS superconductors without any extensions of the model suggested in previous works. We show that the contribution of diagrams with current vertices start to dominate over the density-modulation related processes  for the level of disorder above the extremely weak superclean threshold. 
    Since the superclean regime is hardly realizable in experiments our results provide the basis for the analysis of non-linear responses in realistic superconducting samples to describe the pump-probe or the 
    THG generation experiments  in the broad range of frequencies. The same conclusion holds for compounds with  unconventional pairing such as the  d-wave cuprates \cite{PhysRevLett.120.117001} or multiband superconductors\cite{0295-5075-101-1-17002}.
 Theory suggested in the present paper can be applied to analyse the recent data on the Higgs mode in a d-wave superconductor\cite{PhysRevLett.120.117001} and the collective mode in MgB$_2$ \cite{Giorgianni2019}. In general the impurity scattering determines collective mode excitation in superconductors with both the s-wave and non-trivial pairings and should modify the Higgs mode spectroscopy approach which has been suggested recently\cite{1712.07989} 
 
 Third, we have demonstrated that in the diffusive regime which is typical for thin-film superconducting samples used for recent THz measurements the  resonant contribution to THG signal
  is determined by the Higgs mode excitation thus providing the natural explanation of recent experiments\cite{Matsunaga1145,PhysRevB.96.020505}. 
The amplitude of this peak  is bounded from above by the Dynes parameter which describes the quasiparticle recombination rate due to the electron-phonon interaction.

 \section{Acknowledgements}
  I am grateful to Lara Benfatto for useful discussions. This work was supported by the Academy of Finland (Project No. 297439). 
   
 \appendix
 
 {
 
  \section{Removing singularities in  Eqs.(\ref{Eq:g2s},\ref{Eq:g3s},\ref{Eq:gADelta})}
 \label{AppSec:s14}
 Eqs. (\ref{Eq:g2s}) for the second- and for the third-order
(\ref{Eq:g3s},\ref{Eq:gADelta}) corrections contain 
 the differences $s_i-s_j$ in the denominators. When analytically continued this difference becomes zero for certain energy. Therefore these equations cannot be used directly for the numerical integration along the real energy axis. 
 In order to make them suitable for numerics they Eqs. should be written in such a way to eliminate singular $(s_i - s_j)$ combinations in the denominators. 

 First, we note that one can simplify the expression for the second-order 
  corrections as follows. Substituting Eq. for the first-order correction
  $g_{1a}$ into the Eqs. (\ref{Eq:g2s},\ref{Eq:g3s},\ref{Eq:Y3s})
   we can rewrite them as follows
 \begin{align} \label{Eq:g2sFin}
 &   Z_s(234) \hat g_{2s} (234) = 
    \\ \nonumber
 & \left( s_3 + \tau_{imp}^{-1} \right) 
   \hat\tau_3\hat g_0(3)\hat\tau_3 
   + s_2 \hat g_0(2) + s_4 \hat g_0(4) -   
    \\ \nonumber
  & \left(s_2 + s_3 + s_4 + \tau_{imp}^{-1} \right)
  \hat g_0(2)\hat\tau_3\hat g_0(3)\hat\tau_3\hat g_0(4)
  \end{align}
  
   \begin{align} \label{Eq:g2aFin}
 & Z_a(234) \hat g_{2a} (234)= 
 \\ \nonumber
 &  \tilde s_3 \hat\tau_3\hat g_0(3) \hat\tau_3 
 + \tilde s_2 \hat g_0(2)    
 + \tilde s_4 \hat g_0(4) - 
 \\ \nonumber 
 & \left( \tilde s_2 + \tilde s_3 + \tilde s_4 \right)
 \hat g_0(2)\hat\tau_3\hat g_0(3)\hat\tau_3\hat g_0(4)
 \end{align}   
 where we denote 
   \begin{align} \label{Eq:Za}
   & Z_a(123)= \left( \tilde s_2 + \tilde s_1 \right)
   \left( \tilde s_3 + \tilde s_2 \right)
   \left( \tilde s_3 + \tilde s_1 \right)
   \\ \label{Eq:Zs}
   & Z_s(123)= \left( \tilde s_2 + \tilde s_1 \right)
   \left( \tilde s_3 + \tilde s_2  \right)
   \left( s_3 + s_1 \right) 
   \end{align}

   With the third-order it requires more effort to get 
   rid of the differences $s_1-s_4$ in denominators. 
   First, let us demonstrate how this can be done for the correction 
   $\hat g_{\Delta A}$ that determine the Higgs mode contribution to 
   the current (\ref{Eq:gADelta}). We rewrite it as follows
   $\hat g_{\Delta A} = \hat g^{(1)}_{\Delta A} + \hat g^{(2)}_{\Delta A}$, where
    \begin{align}    \label{appEq:gADelta1}
 & \hat g^{(1)}_{A\Delta} =
 i\frac{
 \hat g_0(1) \hat Y_{A\Delta}
 -
 \hat Y_{A\Delta} \hat g_0(4)}
 {2(s_1 + s_4 + \tau_{imp}^{-1})} 
 \\  \label{appEq:gADelta2}
 & \hat g^{(2)}_{A\Delta} = 
 i\hat g_0(1)\frac{
 \hat Y_{A\Delta}
 +
 \hat g_0(1)\hat Y_{A\Delta} \hat g_0(4)
 }
 {2(s_1-s_4)} 
 \end{align}
 where $\hat Y_{A\Delta} = \hat Y_{\Delta} - I \hat Y_{A} $. 
 Now we need to treat only the  term $\hat g_{A\Delta2}$.
 In order to do that we note the relation 
 \begin{align}
 \hat g_0(1) \hat g_{A\Delta} + \hat g_{A\Delta} \hat g_0(4)=
 i \frac{Y_{A\Delta} + \hat g_0(1) Y_{A\Delta} \hat g_0(4) }
 {s_1-s_4}
 \end{align}
 Using the commutation relationw (\ref{Eq:NormalizationGDelta},
 \ref{Eq:NormalizationGADelta})
 we obtain then the expression for $\hat g^{(2)}_{A\Delta}$
 without singularity
 \begin{align} \label{Eq:g32ADeltaNoSing}
 \hat g^{(2)}_{A\Delta} = - \hat g_1 [ \hat g_{1a}(12)
 \hat g_\Delta(24) + \hat g_\Delta(13)\hat g_{1a}(34) ]/2
 \end{align}
 
 Now let us apply the same trick to the corrections 
 $\hat g_{3s,3a}$. We write them as the superposition of two parts, e.g. $\hat g_{3s}=\hat g^{(1)}_{3s} + \hat g^{(2)}_{3s} $
 \begin{align} \label{Eq:g31}
 \hat g^{(1)}_{3s} = 
 i \frac{\hat g_0(1)\hat Y_s - \hat Y_s \hat g_0(4)}
 {2(s_1+s_4+\tau_{imp}^{-1})}
 \\ \label{Eq:g32}
 \hat g^{(2)}_{3s} =  i \hat g_0(1) 
 \frac{\hat Y_s - \hat g_0(1) \hat Y_s \hat g_0(4)}
 {2(s_1-s_4)}
 \end{align}
 Using commutation relations (\ref{Eq:NormalizationG1a},\ref{Eq:NormalizationG2s},\ref{Eq:NormalizationG3s}) we rewrite expression for 
 $\hat g^{(2)}_{3s}$ in the following form 
 \begin{align}\label{Eq:g32AANoSing}
 \hat g^{(2)}_{3s} =-\hat g_0(1)[\hat g_{1a}(12)\hat g_{2s}(234) + 
 \hat g_{2s}(123) \hat g_{1a}(34)]
 \end{align}
 that does not have singularities. 
 Expressions for the anisotropic part $\hat g_{3a}=\hat g^{(1)}_{3a} + \hat g^{(2)}_{3a}$ are similar to Eqs.(\ref{Eq:g31}, \ref{Eq:g32AANoSing}) with the change of $\hat g_{2s}$ by $\hat g_{2s}$ and $\hat Y_s$ by $\hat Y_a$ .

 \section{Derivation of the response in diffusive limit}
 \label{App:DirtyLimit}
   
 First, in the dirty limit we can find corrections to the propagators 
 directly from the Usadel equation which is a simplified version of 
 the Eilenberger equation.
 In the frequency domain 
 $$
 \hat g_{2}(\tau_1, \tau_2) = 
 T\sum_\omega \left(\frac{\alpha^2 D}{v_F^2}\right) \hat g_{2}(123) e^{i\omega_1\tau_1-i\omega_3\tau_2} 
 $$
 we get from Eq.(\ref{Eq:UsadelGen},\ref{Eq:MWCI},\ref{Eq:CurrentUsadel})
 \begin{align}
 & s_1\hat{g_0}(1) \hat g_{2} - 
 s_3 \hat g_{2} \hat{g_0}(3) = 
 \\ \nonumber
 & \hat g_0 (1) \hat\tau_3 \hat g_0 (2)\hat \tau_3 
 - \hat \tau_3 \hat g_0 (2)\hat \tau_3\hat g_0 (3) 
 \end{align}
 where we denote again 
 $\omega_1 = \omega + 2\Omega$, $\omega_2 = \omega+\Omega$, 
 $\omega_3 = \omega$, $\omega_4 = \omega -\Omega$.
 The solution can be written as (\ref{Eq:g2Usadel})
 \begin{align} \label{Eq:g2UsadelApp}
 g_{2}(123) = 
 \frac{\hat\tau_3\hat g_0 (2)\hat\tau_3 - \hat g_0 (1)\hat\tau_3 
 \hat g_0 (2) \hat\tau_3 \hat g_0(3)}{s_{1} + s_{3}}
 \end{align}  
 Taking into account the diffusion coefficient $D= v_F^2 \tau_{imp}/3$ 
 the solution for $\hat g_{2}$ coincides with $\hat g_{2s}$ 
 obtained from the general expression (\ref{Eq:g2s}) up to the leading
  term in $\tau^{-1}_{imp}$.     
  
 Within the Usadel theory the current  can be calculated using 
 general expression (\ref{Eq:CurrentUsadel}) to have the form (\ref{Eq:Current3OmegaUsadel},\ref{Eq:CurrentH3OmegaUsadel})
 %%%%%%%%%%%%%%%%%%%%%%  
 \begin{align} \label{Eq:Current3OmegaUsadelApp}
 & j_{AAA} =  \frac{\pi D \sigma\alpha^3}{e v_F^3} T \sum_\omega {\rm Tr}
 \hat\tau_3 [ \hat g_0(1) \hat\tau_3 \hat g_{2}(234) + 
 \\ \nonumber
 & \hat g_0(4)\hat\tau_3 \hat g_{2}(123)  ]
 \\ \label{Eq:CurrentH3OmegaUsadelApp}
  & j_{H} =\frac{\pi D \sigma\alpha^3}{e v_F^3}  T \sum_\omega {\rm Tr}
 \hat\tau_3 [ \hat g_0(1) \hat\tau_3 \hat g_{\Delta}(24) + 
 \\ \nonumber
 & \hat g_0(4)\hat\tau_3 \hat g_{\Delta}(13)  ]
 \end{align}
  %%%%%%%%%%%%%%%%%%%%%%%
 Taking the dirty limit for general Eq.(\ref{Eq:jAAAFin}) that determines 
 the current is more tricky. The leading-order correction in the limit 
 $\tau_{imp}\to 0$ is given by (\ref{Eq:g3s}). First, from the 
 Eq.(\ref{Eq:g1a}) we find for the first-order corrects in the limit 
 $\tau_{imp}\to 0$ given by 
 %%%%%%%%%%%%%%%%%%%%%%%5  
 \begin{align}
 & \hat g_{1a}(12) = i\tau_{imp} [ \hat g_0(1)
 \hat\tau_3\hat g_0(2) - \hat\tau_3]
 \end{align}
 %%%%%%%%%%%%%%%%%%%%%%%%  
 Using this relation and the commutation relations 
 $\hat g_{2s}(234) \hat g_0(4) =  - \hat g_0(2) \hat g_{2s}(234) $ we obtain 
 %
 %\begin{widetext}
 \begin{align} \nonumber
 & 2\hat Y_s\hat g_0(4) =  -2\hat g_0(1) \hat Y_s =  
 \\ \nonumber
 & - \left[ \hat\tau_3 \hat g_0(2) + \hat g_0(1)\hat\tau_3 \right]
 \hat g_{2s}(234)  - 
  \hat g_{2s}(123) \left[ \hat\tau_3 \hat g_0(4) +
 \hat g_0(3)\hat\tau_3\right]  
 \end{align}    
 %\end{widetext}
    
 Thus from Eq.(\ref{Eq:g3s}) and Eqs.(\ref{Eq:g31},\ref{Eq:g32AANoSing}) we get 
 \begin{align} \label{Eq:g31sDirty}
 &  \hat g^{(1)}_{3s}  =   \frac{i\tau_{imp}}{2}
  \{ \left[ \hat g_0(1)\hat\tau_3 + \hat\tau_3\hat g_0(2)
 \right] \hat g_{2s}(234) + 
 \\ \nonumber
 & \hat g_{2s}(123) \left[ \hat\tau_3 \hat g_0(4) + 
 \hat g_0(3)\hat\tau_3 \right]  \}
 \end{align}  
  \begin{align} \label{Eq:g32sDirty}
 &  \hat g^{(2)}_{3s}  =   \frac{i\tau_{imp}}{2}
  \{ \left[ \hat g_0(1)\hat\tau_3 - \hat\tau_3\hat g_0(2)
 \right] \hat g_{2s}(234) + 
 \\ \nonumber
 & \hat g_{2s}(123) \left[ \hat\tau_3 \hat g_0(4) - 
 \hat g_0(3)\hat\tau_3 \right]  \}
 \end{align}  
 We substitute the result (\ref{Eq:g31sDirty},\ref{Eq:g32sDirty}) into the expression for current 
 (\ref{Eq:jAAAFin}) and use the summation 
 $\sum_\omega \hat g_0(2)\hat g_{2s}(234) = 
 - \sum_\omega \hat g_{2s}(123)\hat g_0(3) $ to get the expression for the 
 current which is equal to the Eq.(\ref{Eq:Current3OmegaUsadel}).   
  }
 
 The dirty limit for Higgs mode-related part of the current can be 
 obtained from Eq.(\ref{Eq:jHFin},\ref{Eq:g1Delta},\ref{Eq:gADelta},\ref{Eq:YDelta}) in the similar way as above. 
 Using the relation   $\hat g_{\Delta}(24) \hat g_0(4) =  - \hat g_0(2) \hat g_{\Delta}(24) $ 
 we obtain
 \begin{align} \nonumber
 & 2\hat Y_\Delta \hat g_0(4) =  -2 \hat g_0(1) \hat Y_\Delta =  
 \\
 & - [\hat\tau_3 \hat g_0(2) + \hat g_0(1)\hat\tau_3 ]
 \hat g_\Delta (24) 
 -
  \hat g_\Delta (13)
  [\hat\tau_3 \hat g_0(4) + \hat g_0(3)\hat\tau_3 ]
 \end{align}    
  Thus from Eq.(\ref{Eq:gADelta}) and 
  Eqs.(\ref{appEq:gADelta1}, \ref{Eq:g32ADeltaNoSing}) we get 
 \begin{align} \label{Eq:gADelta1Dirty}
 & \hat g^{(1)}_{A\Delta} =
 \frac{i\tau_{imp}}{2}
 \{ \left[ \hat g_0(1)\hat\tau_3 + \hat\tau_3\hat g_0(2)
 \right] \hat g_{\Delta}(24) + 
 \\ \nonumber
 & \hat g_{\Delta}(13) \left[ \hat\tau_3 \hat g_0(4) + 
 \hat g_0(3)\hat\tau_3 \right]  \}
 \end{align}
 \begin{align} \label{Eq:gADelta2Dirty}
 & \hat g^{(2)}_{A\Delta} =
 \frac{i\tau_{imp}}{2}
 \{ \left[ \hat g_0(1)\hat\tau_3 - \hat\tau_3\hat g_0(2)
 \right] \hat g_{\Delta}(24) + 
 \\ \nonumber
 & \hat g_{\Delta}(13) 
 \left[ \hat\tau_3 \hat g_0(4) - \hat g_0(3)\hat\tau_3 \right]  \}
 \end{align}  
 Using the summation 
  $\sum_\omega \hat g_0(2)\hat g_{\Delta}(24) = 
 - \sum_\omega \hat g_{\Delta}(13)\hat g_0(3) $ 
 and the expression for current (\ref{Eq:jHFin})
 we get the dirty limit expression for the current (\ref{Eq:CurrentH3OmegaUsadelApp}).

   \section{Absence of the Higgs mode generation without impurities}
   \label{SecApp:NoDisorderNoHiggs}
   Without disorder we get from Eq.(\ref{Eq:g2sFin}) the 
   expression which according to Eq.(\ref{Eq:FDelta0}) determines the order parameter 
   (\ref{Eq:g2sFin})
    \begin{align} \label{Eq:g2sClean}
   & {\rm Tr} [\hat\tau_2 \hat g_{2s} (123) ]=  
   \\ \nonumber
   &- \frac{\Delta}{s_1s_2s_3 (s_1+s_2)(s_1+s_3)(s_2+s_3) } \times
   \\ \nonumber
   &\left[  
   \left(s_1 + s_2 + s_3  \right)
   (\Delta^2-\omega_1\omega_3 -\omega_2\omega_3-\omega_1\omega_2 )
   + s_1s_2s_3 
  \right]    
  \end{align} 
   Using the relations 
     \begin{align}
     & (s_1+s_2+s_3) (s_1-s_3)(s_1-s_2)(s_2-s_3)=
     \\ \nonumber
    & s_1s_2(\omega_1^2 - \omega_2^2) + s_2s_3(\omega_2^2 - \omega_3^2) 
     + s_1s_3(\omega_3^2 - \omega_1^2) ,
     \\ \nonumber
     & \Delta^2-\omega_1\omega_3 -\omega_2\omega_3-\omega_1\omega_2 = 
     \\ \nonumber
     & s_1^2 - (\omega_1 + \omega_2)(\omega_1 + \omega_3) = 
     s_2^2 - (\omega_1 + \omega_2)(\omega_2 + \omega_3) =
     \\ \nonumber
     & s_3^2 - (\omega_1 + \omega_3)(\omega_2 + \omega_3) 
     \end{align}
 we get 
 \begin{align}
 & \frac{(s_1+s_2+s_3)(\Delta^2-\omega_1\omega_3 -
 \omega_2\omega_3-\omega_1\omega_2 )}
 {s_1s_2s_3(s_1 + s_2 )(s_1 + s_3 )(s_2 + s_3 )} 
  =     
 \\ \nonumber
 & \frac{1}{s_2(\omega_1-\omega_2)(\omega_2-\omega_3)} -
 \frac{1}{s_1(\omega_1-\omega_3)(\omega_1-\omega_2)} 
 -
 \\ \nonumber
 & \frac{1}{s_3(\omega_1-\omega_3)(\omega_2-\omega_3)}      
 +
 \\  \nonumber
 & \frac{s_3}{(\omega_1^2 - \omega_3^2)(\omega_2^2 - \omega_3^2)} 
 +
 \frac{s_1}{(\omega_1^2 - \omega_3^2)(\omega_1^2 - \omega_2^2)}
 - 
 \\ \nonumber
 & \frac{s_2}{(\omega_1^2 - \omega_2^2)(\omega_2^2 - \omega_3^2)} 
 =
 \\
 & \frac{1}{2\Omega^2} \left( \frac{2}{s_2} - 
 \frac{1}{s_1} -  \frac{1}{s_3}\right) - 
 \frac{1}{(s_1+s_2)(s_1+s_3)(s_2+s_3)}
 \end{align}
     Substituting this result into Eq.(\ref{Eq:g2sClean})
 we get as required by the Eq.(\ref{Eq:Trg2Clean})
 \begin{align}
  & {\rm Tr} [\hat\tau_2 \hat g_{2s} (123) ]= 
  -\frac{\Delta}{2\Omega^2} \left( \frac{2}{s_2} - 
  \frac{1}{s_1} -  \frac{1}{s_3}\right) .
 \end{align}
 As explained in the main text since the summation by Matsubara frequencies of this expression yields zero, it yields no perturbation of the order parameter amplitude.      
   
   \bibliography{refsHiggs}
% \begin{thebibliography}{99}
        
% \bibitem{FalkovskiiSuper}
% {\it Inelastic electronic light scattering in superconducting and normal metals with
% impurities}, L. A. Falkovsky, Zh. Eksp. Teor. Fiz. {\bf 103}, 666 (1993) [JETP, {\bf 76}, 331 (1993)].
% 
% \bibitem{AGD}
% A. A. Abrikosov, L. P. Gor'kov, and I. E. Dzyaloshinskii, {\it Quantum
%Field Theoretical Methods in Statistical Physics}, Nauka, Moscow ( 1962)
%[Engl. trans]. Pergamon, Oxford ( 1965) 

% \end{thebibliography}

 \end{document}